\documentclass[10pt,preprint]{aastex}

\usepackage{graphics}
\usepackage{graphicx}
\usepackage{epsfig}
\usepackage{longtable}
\usepackage{blindtext}
\usepackage{float}

\usepackage{color}
\usepackage{amssymb}
\definecolor{red}{rgb}{1,0,0}
\definecolor{black}{rgb}{0,0,0}

\shorttitle{Pluto's atmosphere from stellar occultations}
\shortauthors{Dias-Oliveira et al.}

\begin{document}


\title{Pluto's atmosphere from stellar occultations in  2012 and 2013\footnote{%
Partly based on observations made with the ESO camera NACO at the Very Large Telescope (Paranal),
under programme ID's 089.C-0314(C) and 291.C-5016. The prediction use observations made with the WFI camera at the 2.2 m Telescope,
under programme ID's 079.A-9202(A).
}%
}


\author{Dias-Oliveira, A.\altaffilmark{1,2}, Sicardy, B.\altaffilmark{2}, Lellouch, E.\altaffilmark{2}, Vieira-Martins, R.\altaffilmark{1,3}, 
Assafin, M.\altaffilmark{4}, Camargo, J. I. B.\altaffilmark{1}, Braga-Ribas,~F.\altaffilmark{1,5}, Gomes-J\'unior, A. R.\altaffilmark{4}, Benedetti-Rossi, G.\altaffilmark{1}, 
Colas, F.\altaffilmark{3}, Decock, A.\altaffilmark{6,2}, Doressoundiram, A.\altaffilmark{2}, Dumas, C.\altaffilmark{7}, Emilio, M.\altaffilmark{8},
Fabrega Polleri,~J.\altaffilmark{9}, Gil-Hutton, R.\altaffilmark{11,12}, Gillon, M.\altaffilmark{6}, Girard, J.\altaffilmark{7},  
Hau, G. K. T.\altaffilmark{7}, Ivanov, V. D.\altaffilmark{13,14},  Jehin, E.\altaffilmark{6}, Lecacheux, J.\altaffilmark{2}, Leiva, R.\altaffilmark{15},
Lopez-Sisterna, C.\altaffilmark{11}, Mancini, L.\altaffilmark{16,17}, Manfroid, J.\altaffilmark{6}, Maury, A.\altaffilmark{10}, Meza, E.\altaffilmark{2,18}, Morales, N.\altaffilmark{19},
Nagy, L.\altaffilmark{10}, Opitom, C.\altaffilmark{6}, Ortiz, J. L.\altaffilmark{19}, Pollock, J.\altaffilmark{20}, 
 Roques, F.\altaffilmark{2}, Snodgrass, C.\altaffilmark{21}, Soulier, J.F.\altaffilmark{22}, Thirouin, A.\altaffilmark{23}, Vanzi, L.\altaffilmark{15}, Widemann, 
T.\altaffilmark{2}, Reichart, D. E.\altaffilmark{23},
LaCluyze, A. P.\altaffilmark{24}, Haislip, J. B.\altaffilmark{24}, Ivarsen, K. M.\altaffilmark{24}, Dominik, M.\altaffilmark{25}, 
J\o rgensen, U.\altaffilmark{26}, Skottfelt, J.\altaffilmark{26,27}}

\email{alexoliveira@on.br}

\altaffiltext{1}{Observat\'orio Nacional/MCTI, Rua General Jos\'e Cristino 77, Rio de Janeiro - RJ, 20.921-400, Brazil}
\altaffiltext{2}{LESIA/Observatoire de Paris, CNRS UMR 8109, Universit\'e Pierre et Marie Curie, Universit\'e Paris-Diderot, 5 place Jules Janssen, F-92195 Meudon C\'edex, France.}
\altaffiltext{3}{IMCCE/Observatorie de Paris, 77 Avenue Denfert Rochereau, Paris, 75014, France}
\altaffiltext{4}{Observat\'orio do Valongo/UFRJ, Ladeira Pedro Antonio 43, Rio de Janeiro - RJ, 20080-090, Brazil}
\altaffiltext{5}{Federal University of Technology - Paran\'a (UTFPR / DAFIS), R. Sete de Setembro 3165, Curitiba - PR, 80230-901, Brazil}
\altaffiltext{6}{Institut d'Astrophysique, de G\'eophysique et Oc\'eanographie, Universit\'e de Li\`ege, All\'ee du 6 ao\^ut 17, 4000 Li\'ege, Belgium }
\altaffiltext{7}{ESO, Alonso de Cordova 3107, Vitacura, 7630355 Santiago, Chile}
\altaffiltext{8}{Observat\'orio Astron\^omico Departamento de Geoci\^encias, Universidade Estadual de Ponta Grossa, Paran\'a, Brazil}
\altaffiltext{9}{Observatorio Paname\~no en San Pedro de Atacama, Chile (OPSPA)}
\altaffiltext{10}{San Pedro de Atacama Celestial Explorations (S.P.A.C.E.), San Pedro de Atacama, Chile}
\altaffiltext{11}{Complejo Astron\'omico El Leoncito (CASLEO-CONICET),Av. Espa\~na 1512 sur, J5402DSP San Juan, Argentina}
\altaffiltext{12}{Universidad Nacional de San Juan, Complejo Universitario "Islas Malvinas", Av. Ignacio de la Roza 590 oeste, J5402DCS - San Juan, Argentina}
\altaffiltext{13}{ESO, Karl-Schwarzschild-Str. 2, 85748 Garching bei, M\"unchen, Germany}
\altaffiltext{14}{ESO, Alonso de Cordova 3107, Casilla 19001, Santiago 19, Chile}
\altaffiltext{15}{Instituto de Astrof\'isica, Facultad de F\'isica, Pontificia Universidad Cat\'olica de Chile, Av. Vicu\~na Mackenna 4860, Santiago 7820436, Chile}
\altaffiltext{16}{Max Planck Institute for Astronomy,  K\"onigstuhl 17, D-69117, Heidelberg, Germany}
\altaffiltext{17}{INAF - Osservatorio Astrof\'isico di Torino, via Osservatorio 20, 10025 - Pino Torinese, Italy}
\altaffiltext{18}{Grupo Astronomía, Facultad de Ciencias, Universidad Nacional de Ingiener\'ia, Lima, Peru}
\altaffiltext{19}{Instituto de Astrof\'isica de Andaluc\'ia-CSIC, Apt 3004, 18080, Granada, Spain}
\altaffiltext{20}{Departament of Physics and Astronomy, Appalachian State University, Boone, NC 28608, USA}
\altaffiltext{21}{Planetary and Space Sciences, Department of Physical Sciences, The Open University, Milton Keynes, MK7 6AA, UK}
\altaffiltext{22}{Association Des \'Etoiles pour Tous, 19 Rue Saint Laurent, Maisoncelles, 77320 Saint Martin du Boschet, France}
\altaffiltext{23}{Lowell Observatory, 1400 W Mars Hill Rd, Flagstaff, Arizona, United States of America.  }
\altaffiltext{24}{Department of Physics and Astronomy, University of North Carolina - Chapel Hill, North Carolina 27599, USA}
\altaffiltext{25}{SUPA, University of St Andrews, School of Physics \& Astronomy, North Haugh, St Andrews, KY16 9SS, UK}
\altaffiltext{26}{Niels Bohr Institute, University of Copenhagen, Juliane Maries vej 30, 2100 Copenhagen \O, Denmark}
\altaffiltext{27}{Centre for Electronic Imaging, Department of Physical Sciences, The Open University, Milton Keynes, MK7 6AA, UK}






\begin{abstract}

We analyze two multi-chord stellar occultations by Pluto observed on July 18th, 2012 and May 4th, 2013, and
monitored respectively from five and six sites.
They provide a total of fifteen light-curves, 
twelve of them being used for a simultaneous fit that uses a unique temperature profile,
assuming a clear (no-haze) and pure $N_2$ atmosphere, but allowing for
a possible pressure variation between the two dates. We find a solution that fits satisfactorily (i.e. within the noise level)
all the twelve light-curves, providing atmospheric constraints between
$\sim$1,190~km (pressure $\sim$11~$\mu$bar) and
$\sim$1,450~km (pressure $\sim$0.1~$\mu$bar) from Pluto's center.
Our main results are: 
(1) the best-fitting temperature profile shows a stratosphere with strong positive gradient between 
1,190~km (at 36~K, 11~$\mu$bar) and $r=1,215$~km (6.0~$\mu$bar), 
where a temperature maximum  of 110~K is reached; above it is a mesosphere with negative thermal gradient of -0.2~K~km$^{-1}$ up to 
$\sim$1,390~km (0.25~$\mu$bar), where, the mesosphere connects itself to a more isothermal upper branch around 81~K; (2) the pressure shows a small (6\%) but significant 
increase (6-$\sigma$ level) between the two dates; (3) without a troposphere, Pluto's radius is found to be $R_P = 1,190\pm$5~km.  
Allowing for a troposphere, $R_P$ is constrained to lie between 1,168 and 1,195~km; (4) the currently measured CO abundance is too small
 to explain the mesospheric  negative thermal gradient.
Cooling by HCN is possible, but only if this species is largely saturated; Alternative explanations like zonal winds or vertical
 compositional variations of the atmosphere are unable to explain the observed mesospheric trend.

\end{abstract}





\section{Introduction}

Stellar occultations are a very powerful tool to discover and study, among others, tenuous atmospheres around remote bodies. 
Pluto's atmosphere was discovered using this technique \citep{bro95,ell89,hub88}, and
its spectacular two-fold expansion between 1988 and 2003 was also revealed using stellar occultations \citep{ell03,sic03}.
Other trans-neptunian objects were explored with this technique, and so far, none of
them exhibited atmospheres at the 10 nbar pressure level (three orders of magnitude smaller than for Pluto).
This includes Charon \citep{sic06}, Eris \citep{sic11}, Makemake \citep{ort12} and Quaoar \citep{bra13}.

All those bodies have sizes and surface gravities that are comparable to those of Pluto, within a factor of two.
As such, the derived upper limits constrain 
the physical conditions necessary for the appearance and maintenance of atmospheres around 
a body with a given ice composition and heliocentric distance. 

Here we analyze results derived from two Pluto stellar occultations (18 July 2012 and 04 May 2013) 
that provide signal-to-noise ratios (SNR's) that are among the best ever obtained during such events.
They are furthermore combined with well-sampled multi-chord coverages, providing a good absolute radial scale
for the atmosphere extension.

We use the simplest possible model, assuming a spherically symmetric, clear (no-haze), pure $N_2$ atmosphere
with constant temperature both horizontally and with time. 
Our model satisfactorily fits twelve of the selected light-curves, and provides accurate density, pressure and temperature
profiles for radii between 1,190~km (11 $\mu$bar pressure level) and 1,450~km ($\sim 0.1$ $\mu$bar) from Pluto's center,
while also providing constraints on Pluto's radius.

As Pluto's atmospheric pressure is dominated by the  vapor equilibrium pressure at its surface, 
it is very sensitive to tiny changes of temperature and the available
amount of exposed ice.
This induces strong seasonal effects over the plutonian year \citep{han96}
that can be monitored and analyzed through stellar occultations \citep{you13}. 
In that context, our data reveal a small, but significant increase of pressure between 2012 and 2013, 
which can be used for constraining current Pluto seasonal models, 
see \cite{olk15} for a detailed analysis.

Our results are obtained in the context of the forthcoming flyby of the dwarf planet by the
NASA {\it New Horizons} spacecraft in July 2015.
Consequently, they can be used as a basis of comparison with the {\it New Horizons} findings.




\section{The 2012 and 2013 Pluto stellar occultations}

\subsection{Predictions}

From astrometric observations along Pluto's path onto the sky plane between 2008-2015, 
performed at the ESO's 2.2 m  telescope, \citet{ass10} made accurate predictions for stellar occultations 
involving the dwarf planet and its satellites.

In this context, the two occultations analyzed here, one on 18 July 2012 and the other on 04 May 2013,  
stood out as promising events, owing to the magnitudes of the candidate stars and 
to the presence of several potential observing sites along  the shadow's path.

Follow-up astrometric observations of the stars were carried out in order to improve the predictions. These observations 
were made with the 1.6 m (Perkin-Elmer) and 0.6 m (Boller \& Chivens) telescopes, at Pico dos dias Observatory (OPD, IAU code 874),
and they are done wherever possible within our access time. 

Moreover, 16 positive detections of other occultations by Pluto, that occur between 2005 and 2013, were used to improve
Pluto's ephemeris offset (see \citet{ben14} for details).  

Days before the event, we carried out observations with Pluto and the occulted star present in the same field of view 
of our CCD's in order to minimize systematic biases like catalog errors.

\subsection{Observations}

\begin{figure*}[!htb]
  \centering
\includegraphics[height=6cm,trim= 100 80 100 80]{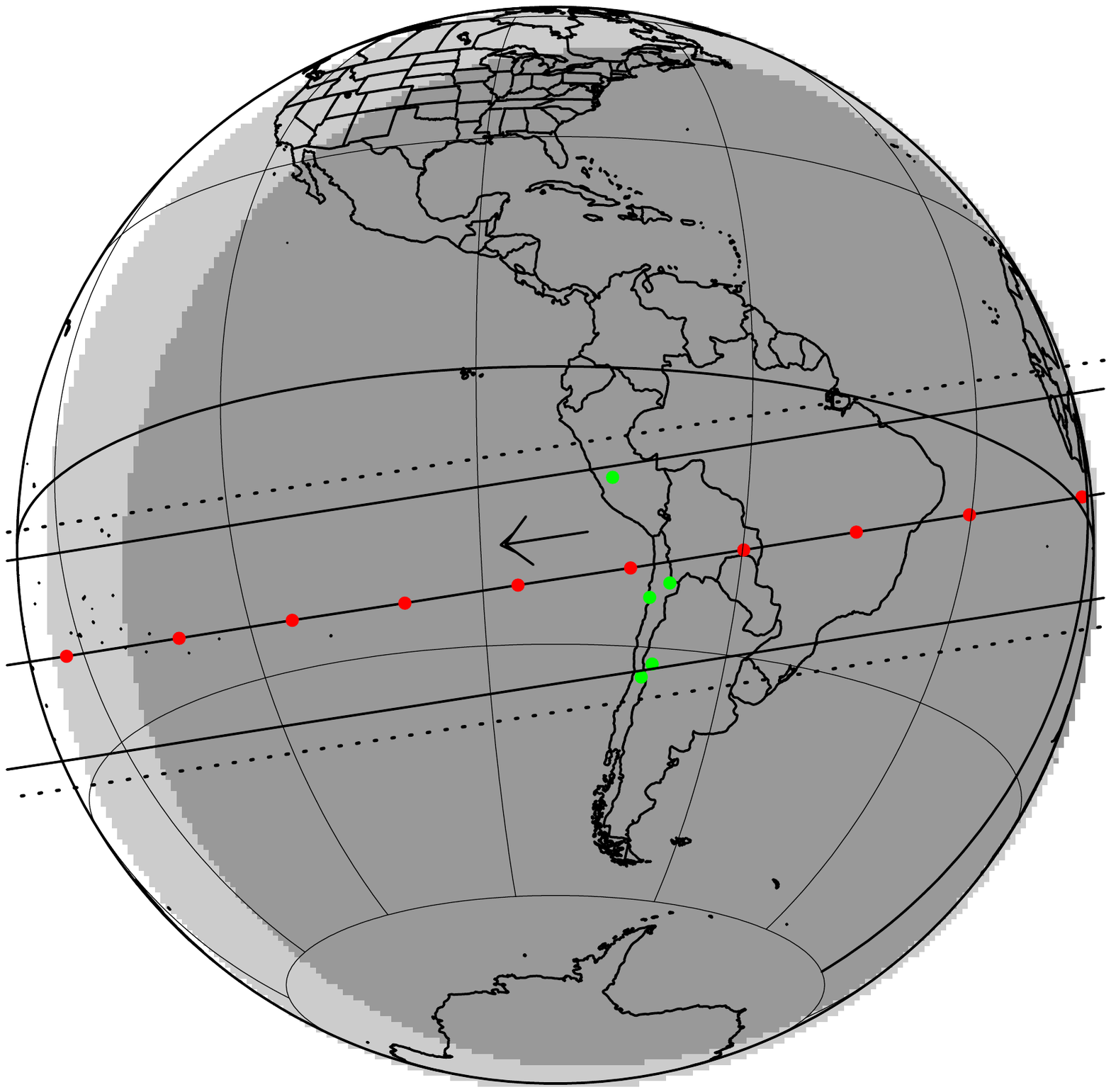}
\includegraphics[height=6cm,trim= 100 80 100 80]{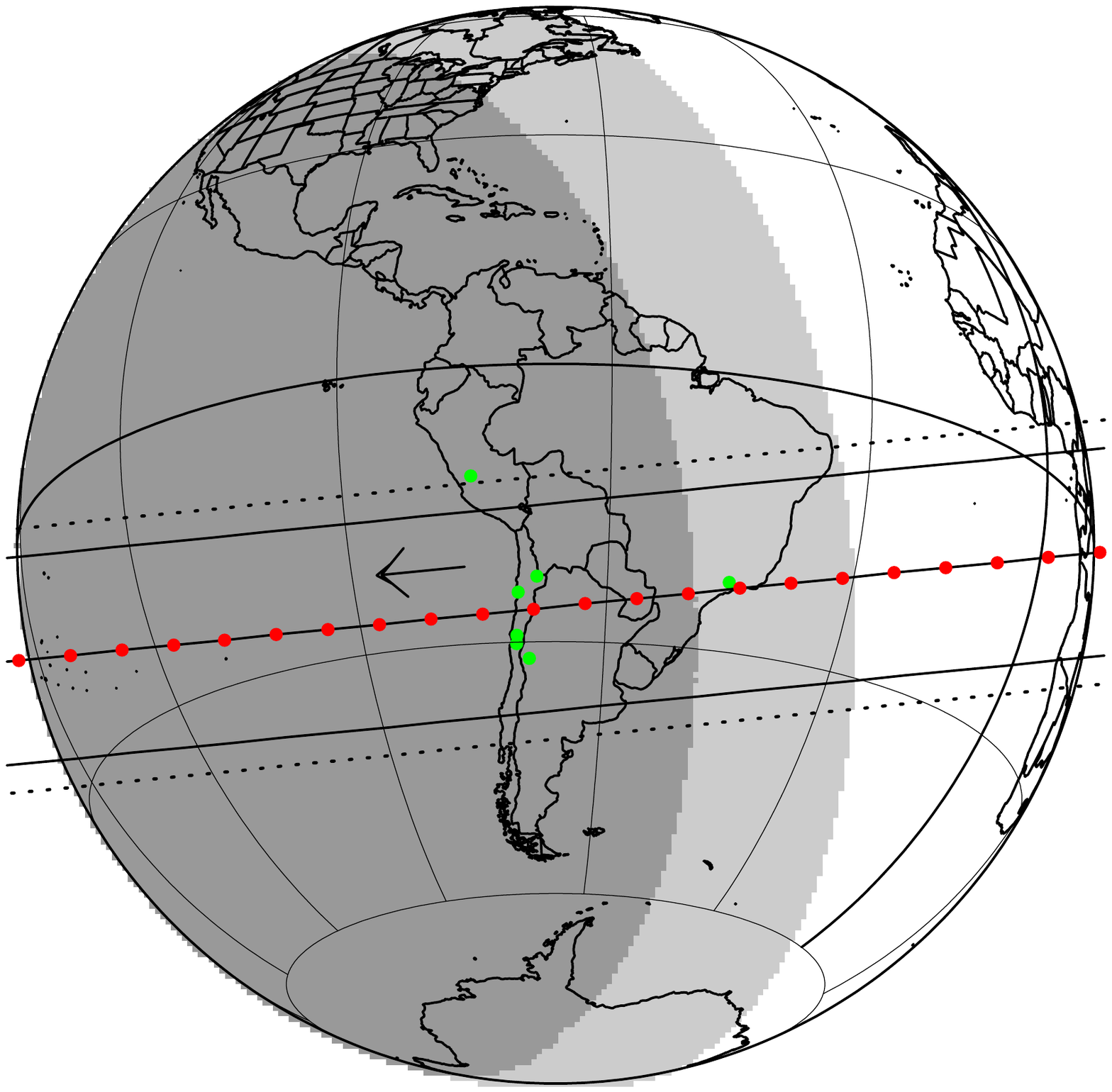}
\caption{%
\normalsize
The post-occultation, reconstructed paths of Pluto's shadow for the two events studied here.
The red dots indicate the shadow center every minute and 
the arrows show the direction of motion.
The green dots mark the sites where data were obtained.
The black solid lines correspond to the half-light stellar level, 
while the dotted lines correspond to the 1\% stellar drop, thus marking the 
practical region of detectability of the occultations.
Left - The 18 July 2012 event. The first red dot at right is at 04h 09m UT, the last one at left corresponds to 04h 18m UT.
Right~-The 4 May 2013 event. The first red dot at right is at 08h 12m UT, the last one at left corresponds to 08h 33m UT.
}%
\label{fig_map}
\end{figure*}

The 18 July 2012 Pluto occultation was observed near zenith from five sites in South America (Fig.~\ref{fig_map}).
The circumstances and technical details of the observations are provided in Table~\ref{tab_18jul}.
The 04 May 2013 event was recorded from six sites, under similar conditions (Fig.~\ref{fig_map}),
providing ten light-curves (Table ~\ref{tab_04may}).
Various astrometric, photometric and physical parameters associated with 
each event are summarized in Table~\ref{tab_param}.

Fig.~\ref{fig_XY} displays  the reconstructed geometries of each event,
showing the Plutocentric latitudes and altitudes probed by the primary stellar image at each site,
see appendix for details. 
For Paranal (18 July 2012), we plot for sake of illustration the trajectories of 
both  primary and secondary stellar images.
As commented later, 
the contribution of the secondary image is small but not negligible compared to that of the primary image
near the shadow center.
Note that the primary image probes the summer (resp. winter) Pluto's hemisphere at ingress (resp. egress).

\subsection{Photometry and calibration}

Classical bias, dark, flat-field and sky subtraction provide the occultation light-curves displayed in 
Figs.~\ref{fig_fit_18jul12}, \ref{fig_fit_04may13} and \ref{fig_leo_dan_spa}.
In all cases, a reference star brighter than the target  was used to correct for low-frequency
transparency variations. 

As expected, the best  SNR light-curve was obtained at Paranal on 18 July 2012,
using the NAOS-CONICA\footnote{NAOS-CONICA is Nasmyth Adaptive Optics System (NAOS) and Near-Infrared Imager and Spectrograph (CONICA)} (NACO) \citep{lez03,rou03} camera attached to the 8.2-m ``Yepun" Very Large Telescope (VLT)
of the European Southern Observatory (ESO),
at a rate of 5 frames per second in H band.
Moreover, this is the only data set for which we have an accurate photometric calibration, 
which allows us to subtract the contribution of Pluto and Charon from the occultation light-curve,
see below.
As such, the 18 July 2012 data provide the best constraints on Pluto's atmospheric structure.
However, the 04 May 2013 light-curves have on average better SNR's than those of 18 July 2012, 
as well as a better spatial sampling, 
thus providing better constraints on the absolute vertical scale of the atmosphere.

Calibration images were taken with NACO some twenty minutes before the 2012 event.
They show resolved images of Pluto, Charon and  the star under excellent seeing conditions (Fig.~\ref{fig_coro}).
Digital coronagraphy \citep{ass08,ass09} was used to remove the star contamination from Pluto and Charon images.
Classical aperture photometry finally provided the Pluto + Charon flux relative the occulted star.
This allows us to estimate the residual stellar fux in the deepest part of the 18 July 2012 occultation at Paranal,
with a value that varied from $2.3 \pm 0.8$\% to $1.8 \pm 0.8$\% of its unocculted value in the central part
of the occultation (Fig.~\ref{fig_fit_18jul12_NACO}).

\section{Modeling of Pluto's atmosphere}

The general idea for modeling Pluto's atmosphere is to use an iterative procedure,
combining both direct ray-tracing and inversion approaches.
We first invert our best signal-to-noise ratio light-curve to retrieve
Pluto's atmospheric density, pressure and temperature profiles (see Appendix and \cite{vap73}).
The retrieved temperature profile is then used as a guide to generate, through direct ray-tracing, 
synthetic occultation light-curves that are simultaneously fitted to all the observed light-curves 
obtained at a given date. 
This pins down the location of Pluto's shadow center relative to the occultations chords
for both the 2012 and 2013 events (Fig.~\ref{fig_XY}).
Finally, the inversion of the best light-curve is performed again and the procedure is resumed. 
This iterative process eventually provides the accurate geometry of each event, 
as well as consistent density, pressure and temperature profiles that best fit
all the occultation light-curves.

Simplifying assumptions are made in our procedure (possible caveats are discussed later): 
$(i)$
Pluto and its atmosphere are spherically  symmetric, 
all quantities depend only on the radius $r$ (defined as the distance to Pluto's body center), 
$(ii)$
the atmosphere is transparent (no haze present), and
$(iii)$
it is an ideal gas in hydrostatic equilibrium, 
in our case, a pure molecular nitrogen $N_2$ atmosphere, 
neglecting other minor species like methane. Moreover,
$(iv)$
we assume that $T(r)$ is time-independent, i.e. the temperature profiles are the same in 2012 and 2013.
Once $T(r)$ is derived as  detailed later, the density and pressure profiles $n(r)$ and $p(r)$ are derived from the 
hydrostatic and ideal gas equations (Eq.~\ref{eq_n_r}), 
once a boundary condition is provided, i.e. the pressure at a given radius.
$(v)$
Although $T(r)$ is taken as time independent, 
the pressure is not.
This is justified by the fact that the pressure is very sensitive to Pluto's surface temperature 
through the vapor pressure equilibrium equation. For instance a 1-K temperature increase 	at the 
surface results in a two-fold increase of pressure or so (Fig.~\ref{fig_T_r_invers_par12}).
Thus, the pressure is a free parameter in our fits. 
More precisely, Eq.~\ref{eq_n_r} requires a boundary condition, once $T(r)$ is fixed.
So, we use as a free parameter the pressure $p_r$ at an arbitrary radius $r$. 
We choose $r=1,275$~km for an easier comparison with other works 
that provide the pressure at  that level (see e.g. \cite{olk15}). 
This level corresponds to a normalized stellar flux of $\approx 0.45$ in the shadow plane.
Once $p_{1,275}$ is given,  the density and pressure profiles $n(r)$ and $p(r)$ are uniquely defined.

We choose the 18 July 2012 occultation light-curve obtained at VLT/NACO to perform the first inversion. 
We use this particular light-curve because
it has the highest SNR of all (Fig.~\ref{fig_fit_18jul12}), and also because 
this is the only one for which we have a reliable measurement of the background contribution from Pluto and Charon 
(Fig.~\ref{fig_fit_18jul12_NACO}), necessary to correctly invert any occultation light-curve.
The successive steps of our procedure are as follows: 

(1) The inversion reveals a strong increase of temperature just above the surface (stratosphere), 
followed by a turning point where the temperature reaches a maximum (stratopause), 
then  a region with a mild negative gradient (mesosphere), and finally 
an isothermal upper branch, see Appendix and Fig.~\ref{fig_model}.
Using the prescriptions described by Eqs.~\ref{eq_T_r_model}, we adjust the coefficients $c_1,...c_9$
controling the profile $T(r)$ in order to best fit the inverted temperature profiles (see Table~\ref{tab_atmo_fit}).

(2) Keeping the profile $T(r)$ fixed in shape, we simultaneously fit 
seven of the light-curves obtained on 04 May 2013.
The free parameters of that fit are the two coordinates defining the shadow center,
the  pressure $p_{1,275}$ at radius 1,275~km, and 
the value of $r_1$, the deepest point that we consider in our profile.
At this stage, when $r_1$ is varied, all the other radii $r_2$, $r_3$ and $r_4$ defining $T(r)$,
see Eq.~\ref{eq_T_r_model}, 
are changed by the same amount. In other words, the entire profile $T(r)$ is vertically displaced by this amount.
Thus, $r_1$ eventually fixes the absolute vertical scale of the atmospheric profile.
Note that $r_1$ is {\it not}, a priori, the radius of the stratobase, nor Pluto's surface radius.
In practice, the choice of $r_1$ is made so that the stellar rays from the faint secondary image passing at $r_1$
have a contribution to the total flux that is negligible compared to the light-curve noise level. 
Thus, taking larger values of $r_1$ would create artificial discontinuities in the synthetic light-curve,
while smaller values would require useless computation time.
To find Pluto's shadow center, we separate the fit along the direction of
the star motion relative to Pluto from the fit perpendicular to that direction.
This is because the fit along the star motion is essentially independent of
the atmospheric model, and is generally more accurate than the fit perpendicular
to that direction.

Note that the 2013 light-curves have generally a better SNR than those of 2012 
(excluding the VLT data set),
because of a better distribution of the chords (Fig.~\ref{fig_XY}).
Consequently, the 2013 occultation light-curves provide a better constraint for $r_1$,
or equivalently, for the absolute vertical scale of the atmospheric model, than those of 2012.

(3) Fixing $r_1$ to its value found in step (2), 
we turn back to the 18 July 2012 data set and simultaneously
fit the five corresponding light-curves, varying Pluto's shadow center and $p_{1,275}$. 

The procedure is then resumed at point (1). It is a converging process
that provides consistent solutions for 
the shape of the profile $T(r)$, 
the absolute vertical scale for $T(r)$,
the centers of Pluto's shadow for both events, and 
the two boundary conditions $p_{1,275}$ for the 18 July 2012 and 04 May 2013 events.
This fitting procedure has a total of twelve free parameters: 
the nine coefficients $c_1,...c_9$, 
the two coordinates that define Pluto's shadow center, and
the pressure $p_{1,275}$.

As commented before,
the 18 July 2012  NACO light-curve is the only one for which the Pluto + Charon contribution is measured
(Fig.~\ref{fig_fit_18jul12_NACO}).
So, the stellar flux was normalized between that value and the full unocculted flux before starting
the fit procedure.

For the other light-curves, the inverse approach was used: 
the background Pluto + Charon flux was imposed
by linearly adjusting the normalized, synthetic stellar flux to the
actual occultation light-curve, through a least-square fit. 
As the residual stellar flux well inside the shadow is mainly
controlled by the density scale-height of the deep stratosphere (Eq.~\ref{eq_H_bottom}), 
this means that the structure of that region is in fact dominated by the NACO, 18 July 2012 data. 
The other light-curves thus mainly serve to constrain the atmospheric structure above that level (mesosphere).  

\section{General atmospheric structure}

The best fits of our synthetic light-curves to the data are shown in Figs.~\ref{fig_fit_18jul12} and \ref{fig_fit_04may13}.
For each light-curve, the residuals are displayed at the bottom of the corresponding panel. 
They show that a unique global model satisfactorily explains all the observations, with 
$\chi^2$ values per degree of freedom ($\chi^2_{\rm dof}$, see Eq.~\ref{eq_chi2}) 
close to unity, except for the 18 July 2012 NACO data (Figs.~\ref{fig_fit_18jul12} and \ref{fig_fit_04may13}).
In fact, due to the quality of this particular data set, the residuals are dominated by 
spikes associated with wave activity, as illustrated in Fig.~\ref{fig_fit_18jul12_NACO}.
The wave activity, including the one observed in the NACO data, 
is discussed in details elsewhere, see \cite{fre15}.

The parameters of the best atmospheric model are listed in Table~\ref{tab_atmo_fit}.
Note that the only parameter that differs between 18 July 2012 and 04 May 2013 is 
the boundary condition, i.e. the pressure $p_{1,275}$ at $r=1,275$~km.
Table~\ref{tab_atmo_fit} reveals a small (6\%) but significant (6-$\sigma$ level) increase of pressure, from 
$p_{1,275} =2.16 \pm 0.02$ $\mu$bar in July 2012 to  
$p_{1,275}= 2.30 \pm 0.01$ $\mu$bar in May 2013, 
corresponding to a pressure increase rate of 7.5\% per year.

Based on various occultation data collected, \cite{you13} and \cite{olk15} 
report a general pressure increase of some 3.5-7.5\% per year between 2006 and 2013,
consistent with our result above.
Note that our value of $p_{1,275}=2.30 \pm 0.01$ $\mu$bar for May 2013 differs
from \cite{olk15}'s result ($2.70 \pm 0.2$ $\mu$bar) by a barely significant $0.4 \pm 0.2$ $\mu$bar. 
Part of this difference could be due to the different methods used to derive those numbers, 
as \cite{olk15} use an isothermal fit to the upper part of the light-curves, while we use
a combination of mesosphere with negative thermal gradient and an upper isothermal branch (Fig.~\ref{fig_model}).

Fig.~\ref{fig_T_r_invers_par12} displays
the density vs. radius, 
the temperature vs. pressure, 
the temperature vs. radius, and
the temperature gradient of  our best model.
Also shown superimposed in that figure are the ingress and egress profiles retrieved from 
the inversions of the 18 July 2012 NACO light-curve,
corresponding respectively to the summer and winter hemispheres, 
as far as the primary stellar image is concerned.
Fig.~\ref{fig_T_r_invers_par12_bottom}  is a more detailed view of the bottom of the 
temperature and temperature gradient  profiles, close to Pluto's surface.

The shaded areas in Fig.~\ref{fig_T_r_invers_par12} and \ref{fig_T_r_invers_par12_bottom} 
indicate the 1-$\sigma$ error envelopes caused by 
$(i)$ the photometric noise in the NACO light-curve, 
that mainly affects the upper parts of the profiles, and
$(ii)$ the uncertainty on the Pluto + Charon contribution to the total observed flux,  
that mainly affects the lower parts of the profiles.
The methods to calculate these uncertainty domains are described in the Appendix.
The temperature profiles are furthermore affected by another source of uncertainty, namely 
$(iii)$ the a priori unknown temperature  boundary condition, 
inherent to the nature of Eq.~\ref{eq_n_r} (a first order differential equation).
As examples, we show in Fig.~\ref{fig_T_r_invers_par12} (gray lines in panels b, c and d) 
the profiles obtained by changing by $\pm$5~K the nominal boundary condition 
($T=80.5$~K at $r=1,390$~km) of the egress, inverted NACO temperature profile. 

Both the photometric noise and the unknowledge of the temperature boundary condition cause an exponential divergence 
of the uncertainty domain for $T$ and $dT/dr$ as $r$ increases, with an $e$-folding distance equal 
to the density scale height $H$ (Eq.~\ref{eq_sigma_all} and Fig.~\ref{fig_T_r_invers_par12}). 
Nevertheless, we note that {\it if} we have independent information on
Pluto's atmosphere, e.g. from theoretical models or forthcoming observations from the {\it New Horizons} mission, 
then we can constrain our temperature at rather high altitudes. 
For instance, at radius $r=1,450$~km (pressure $\sim$0.1~$\mu$bar), the
1-$\sigma$ uncertainty on $T$ caused by photometric noise is about $\pm$2.5~K.
Conversely, the two alternative solutions  $T(r)$ given as examples in Fig.~\ref{fig_T_r_invers_par12} 
(the gray lines in panel (c)),
using different  boundary conditions, differ from each other by 30~K at that same radius.
Consequently, they can be distinguished well above the noise level
if we dispose of independent constraints on the thermal properties of the atmosphere at that radius.

For instance, the warmer gray profile with strong positive temperature gradient in 
Fig.~\ref{fig_T_r_invers_par12} can be discarded if we adopt current models which predict that UV heating 
is efficient only at much higher levels \citep{zhu14}.
The same is true for the cooler gray temperature profile in panel (c) of Fig.~\ref{fig_T_r_invers_par12}:
it shows too strong a negative gradient in the 1,400-1,450~km range, considering that 
atmospheric escape may cause temperatures as low as $\sim 60$~K, but only much higher in the
atmosphere (Ibid.).

Figs.~\ref{fig_T_r_invers_par12} reveals three regions in our thermal profile, from bottom to top: 
a stratosphere with strong positive gradient that starts around 
1,190~km with temperature near 36~K and pressure 11~$\mu$bar, and 
reaches a maximum temperature of 110~K at the stratopause 
(near $r=1,215$~km, 6.0 $\mu$bar).
Then follows a mesosphere with mild negative thermal gradient of -0.2~K~km$^{-1}$ up to the mesopause 
($r \sim 1,390$~km, 0.25~$\mu$bar), 
where, it connects itself to a more isothermal upper part around 81~K.
These regions are now described in detail.

\section{Stratosphere}

As explained in the Appendix, the residual stellar flux in the mid-part of the occultation 
is proportional to the local density scale-height $H$, which is itself related to the strong stratospheric
temperature gradient (Eq.~\ref{eq_H_bottom}).

It is important to note that at closest approach to Pluto's shadow center on 18 July 2012, 
our model predicts that the secondary image observed at Paranal contributes by 20\% to the total, 
primary + secondary stellar residual flux (Fig.~\ref{fig_fit_18jul12_NACO}). 
This is not negligible and explains why we have to extend our ray tracing model {\it below} 
the deepest  radius obtained for the inverted temperature profiles 
(red and blue lines in Fig.~\ref{fig_T_r_invers_par12_bottom}).
In fact, the inversion procedure assumes that there is only one (primary) stellar image contributing
to the flux at any moment, while the direct ray tracing procedure does account for the presence 
of the two images.
When the secondary image appears and disappears (at the extremities of the orange trajectory
shown in Fig.~\ref{fig_XY}), it reaches the radius $r_1= 1,190.4$~km (Table~\ref{tab_atmo_fit}).
Its appearance and disappearance cause small discontinuities in the synthetic flux, but
they are too small to be distinguished from the noise  (Fig.~\ref{fig_fit_18jul12_NACO}).

Due to the uncertainty on the Pluto + Charon flux contribution, the deepest point of our model
is determined to be at  $1,190\pm5$~km (Fig.~\ref{fig_T_r_invers_par12_bottom}).
At that point, nitrogen  reaches its saturation vapor pressure (Fig.~\ref{fig_T_r_invers_par12}),
and thus condenses in principle into ice, i.e. reaches Pluto's surface.
In that context, we obtain a solution with a clear nitrogen atmosphere and
a Pluto radius of  $1,190\pm5$~km which consistently explains all our observations, 
accounting for the presence of both the primary and secondary images.

Other models are possible, though. 
Based a more incomplete and lower quality data set than used here, \cite{lel09} conclude that the 
nitrogen condensation level occurs somewhere in the range 1,187-1197~km, 
consistent with the present work.
However, a shallow adiabatic troposphere with dry or wet nitrogen (or methane) may exist below 1,190~km.
Nevertheless, there is little freedom for such tropospheric models because 
$(i)$ they tend to create caustics in the light-curves that are not observed and 
$(ii)$ they provide a cold methane column density that would be detected by other means.
More precisely, using spectral data, \cite{lel09} find possible tropospheric solutions in a narrow region 
of the parameter space, with depths that cannot exceed 17~km.
Similarly, combining again constraints from spectra with a preliminary analysis 
of the occultation data presented in this work, \cite{lel15} concluded  that Pluto's radius should be between 1,180-1,188 km,
some 2-8~km below the condensation radius 1,190~km derived above.

This said, we assume here that the atmosphere is haze-free, a subject of debate since the discovery
of Pluto's atmosphere. 
Analyzing a high SNR occultation observed in 2006, \cite{you08} conclude that
a haze-only explanation for the light-curve is extremely unlikely. 
In fact, the clear atmosphere model implies a temperature profile that naturally
connects the maximum temperature of $\sim$110~K near 1,215~km to the surface
at average temperature of $\sim$50~K \citep{lel00,lel13}, see Fig.~\ref{fig_T_r_invers_par12} .

Other constraints come from a central flash observation during a stellar occultation in July 2007.
From that event,  \cite{olk14} conclude that the flash is consistent with a transparent atmosphere
with temperature gradient of 5~K~km$^{-1}$ at 1,196~km, fully consistent with our 
results (Fig.~\ref{fig_T_r_invers_par12_bottom}). 
\cite{olk14}  exclude in particular a haze-only model to explain the central flash, although
combinations of thermal gradient and haze mechanism are possible.
In the same vein, \cite{gul15} use a wavelength-resolved occultation on 2011 to constrain
the presence of hazes in Pluto's atmosphere. Although haze models do improve the
fit residuals, a clear atmosphere with a steep thermal gradient at the bottom is also
consistent with the observations.

Finally, we note that the residual stellar flux exhibits a significant decrease in the bottom of the light-curve, 
from 2.3\% to 1.8\% of its unocculted value, 
in the central part of the occultation as observed from Paranal on 18 July 2012 (Fig.~\ref{fig_fit_18jul12_NACO}).
This behavior was already pointed out by \cite{sic03}, based on another hight SNR occultation
observed in August 2002. 
In both cases, the residual stellar flux decreased as the primary stellar image scanned first
the summer, permanently lit northern lower atmosphere, and then the winter, low insolation
region (Fig.~\ref{fig_XY}).
This point is discussed in the last section.

\section{Mesospheric negative temperature gradient}

Above the stratopause ($r \sim 1,215$~km), 
the temperature profile exhibits a negative temperature gradient
up to $r \sim 1,390$~km, with an average value of $dT/dr \sim -0.2$~K km$^{-1}$.
In this $\sim$~170 km radius interval, the temperature decreases by some 30~K. 
This mesospheric gradient is little affected by the choice of the temperature boundary condition, 
see Fig.~\ref{fig_T_r_invers_par12}, panel (d). 
While the photometric noise and the boundary condition problem induce rapidly diverging solutions for 
$dT/dr$ above $\sim$~1,400~km,  the thermal gradient between 
$\sim$~1,250~km and  $\sim$~1,360~km is robustly constrained around -0.2~K~km$^{-1}$,
with a typical fine-scale scatter of $\pm$0.05 ~K~km$^{-1}$ that is dominated by 
Pluto's wave activity, and not by the photometric noise.
In this interval, the thermal gradient remains smaller (in absolute value) than 
the dry adiabatic lapse rate $-g/c_p$ (Fig.~\ref{fig_T_r_invers_par12}), 
where $g$ is the acceleration of gravity (Eq.~\ref{eq_g_r}) and
$c_p=  1.04 \times 10^{3}$~J K$^{-1}$ kg$^{-1}$ is the specific heat at constant pressure for $N_2$. 
Thus, the mesosphere remains convectively stable.

Note that in principle we may choose an extreme temperature boundary condition that provides an 
isothermal mesosphere, i.e. a thermal profile that is much warmer than the warmer gray profile shown in panel (c)
of Fig.~\ref{fig_T_r_invers_par12}. As commented earlier, however, this replaces one problem by another one,
namely that the upper part of our profile is too warm, with seemingly no plausible physical explanation. 

The negative mesospheric thermal gradient is further confirmed by the inversions of our best SNR light-curves 
obtained in July 2012 and May 2013, see Fig.~\ref{fig_T_r_invers}.
This eliminates random, low frequency sky-transparency variations that may have corrupted the light-curves.
Moreover, such gradients have also been reported in previous, independent works.
For instance  \cite{you08} derive and discuss a $dT/dr= -0.086 \pm 0.033$~K km$^{-1}$ gradient at $r= 1,275$~km 
from the 12 June 2006 occultation, while 
\cite{ell07} give $-0.17 \pm 0.05$~K km$^{-1}$ for the same occultation and at the same radius.
\cite{gul15} report a gradient of $-0.23 \pm 0.05$ K km$^{-1}$ in the 1,310-1,450~km region
from the 23 June 2011 occultation, consistent with \cite{per13} for that event.
Finally, \cite{bos15}  derive values of $-0.17 \pm 0.03$~K km$^{-1}$ and $-0.24 \pm 0.01$~K km$^{-1}$
around 1,280-1,300~km, 
for occultations observed on 09 September 2012 and 04 May 2013, respectively.

%
The origin of this thermal gradient is still debated.
Two classes of possible explanations can be proposed: 
(1) the presence of cooling minor species and 
(2) yet unmodeled physical mechanisms. 
They are now examined in detail.

\subsection{Possible cooling by CO or HCN}

Radiative-conductive models of Pluto's atmosphere have been developed initially by \cite{yel89,hub88,lel94,lel15} 
to explain the recently-discovered gross characteristics of Pluto's atmosphere:  
a large lower atmosphere temperature gradient, and a warmer ($\sim$$100$~K) mesosphere.
These studies used a simplified description of the heating/cooling properties of Pluto's atmosphere proposed by \cite{yel89}, 
with heating in the methane 3.3 $\mu$m band and cooling in its 7.6 $\mu$m band, both occurring in non-LTE conditions. 
\cite{lel94} first suggested that additional cooling due to LTE CO emission rotational lines 
was important, based on an estimated abundance of CO in Pluto's atmosphere ($10^{-4}$-$10^{-3}$).

These studies were updated with the much more extensive model of \cite{str96}. 
Notably these authors improved the treatment of solar heating in the CH$_4$ near-infrared bands by considering 
the effects of opacity and vibrational (V-V and V-T) energy transfer, and showed the
need to include heating from the 2.3 $\mu$m band system in addition to the 3.3 $\mu$m bands.

As the composition of Pluto's atmosphere, as well as surface (pressure, radius) conditions, 
were largely unconstrained at that time, \cite{str96} explored diverse combinations of surface pressure and 
methane mixing ratios (including non-uniform ones), including also the effect of CO cooling. 
In general these models were reasonably successful at explaining large near-surface temperature gradients, 
though
$(i)$ fitting 10-20 K km$^{-1}$ gradients near the surface required pushing the models to their limits,  e.g. 
a 3.6\% CH$_4$ mixing ratio confined to the first scale height near the surface and 
a 3 $\mu$bar surface (or tropopause) pressure; 
$(ii)$ models tended to overestimate the upper atmosphere temperature ($\sim 130$~K instead of 100~K).
A general feature of the \cite{str96} models was their prediction of a mostly isothermal atmosphere
at pressures less than $\sim 2$ $\mu$bar, though some models exhibited a moderate  (0 to -0.1 K km$^{-1}$ 
negative gradient at 1-2 $\mu$bar.
As the direct detection of $N_2$ in Pluto's atmosphere is still missing, they also considered a CO-dominated 
atmosphere case (e.g. 97\% CO + 3\% CH$_4$). 
This case led, through enhanced CO cooling, to much larger negative temperature gradients in the sub-microbar region 
and an upper atmosphere temperature of about 55~K.

The avaibility of new, quantitative, observational constraints on 
the composition  (CH$_4 \sim 0.5$\%, CO $\sim$ 0.05\%) 
and near-surface structure (surface radius and pressure, tropospheric depth) of Pluto's atmosphere from 
near-IR observations \citep{lel09,lel11} prompted a revival of the \cite{str96} models \citep{zal11a,zal11b,zhu14}. 

Model updates included new estimates of the 
vibrational energy transfer based on recent laboratory measurements of collisional relaxation rates \citep{sid94,bou03}, as well 
as the introduction of a scheme parameterizing the processes of eddy mixing and convection. 
With the updated model, \cite{zal11a} explored the effect of parameter space 
(CH$_4$ and CO mixing ratios, surface pressure and radius) allowed by the recent observations, 
assuming uniform vertical mixing of CH4 and CO
(which was recently demonstrated to be the case for CH$_4$ in the first ~25 km of the atmosphere, \cite{lel15}).
Radiative-convective calculations were then coupled to a model generating synthetic occultation 
lightcurves for direct comparison to observations. The study was extended by \cite{zal11b}
to include a putative troposphere. 

In spite of minor changes, the \cite{zal11a,zal11b} models confirm 
the essential features of the earlier \cite{str96} models. 
The stratopause temperature is still somewhat too high (120-125~K) near 1,215~km radius in \cite{zal11a}.
These models generally show only weak negative temperature gradients above this level, 
typically a $\sim$5~K decrease over a 300~km range for a CO mixing ratio of $5 \times 10^{-4}$,
or mild $\sim$10~K decrease due to atmospheric escape \citep{zhu14}.
This is in disagreement with the profile reported in the present study, which exhibits a typical 30~K decrease 
between  1,215-1,390~km and a gradient of -0.2 K km$^{-1}$, as discussed earlier. 
Rather, the profile we derive is remarkably similar to that calculated by \cite{zal11a} (their Fig.~8) 
for the case of 40-times enhanced CO mixing ratio ($200 \times 10^{-4}$) 
This scenario, however,  is at odds with the direct measurement of the CO abundance \citep{lel11}.
This suggests that an additional cooling source is at work. 

Through radiation in its intense rotational lines, HCN is the major cooling agent in Titan's upper atmosphere, 
where its mixing ratio is typically $2 \times10^{-4}$ at 1,100~km \citep{vui07}, and 
where it equilibrates the solar UV heating rates \citep{yel91}. HCN has not been detected yet in Pluto's atmosphere, 
but its presence is expected from coupled photochemistry in a N$_2$-CH$_4$ atmosphere.
A complete re-assessment of the Pluto models is however beyond the scope of the present study. 
Here we only re-calculate CO cooling rates, and also examine the case of HCN cooling.  
Photochemical models predictions lead to rather diverse mixing ratios of HCN (10$^{-8}$-10$^{-3}$ of $N_2$ 
\citep{sum97,lar97,kra99}, where the difference largely seems to come from the fact that the more ``optimistic" models 
have not accounted for the fact that under cold ($< 100$~K) temperatures, 
atmospheric condensation of HCN should occur.  
Here, we nominally  consider cases in which the HCN abundance is limited by the  saturation law \citep{fra09}, 
but we also run a case with uniformely mixed HCN, as supersaturation may be possible in a clear, tenuous 
atmosphere as Pluto's. %
Cooling rates at radius $r$ are calculated from the following equation (e.g. for CO):
\begin{equation}
R_{CO}(r)= 4 \pi N_{CO}(r) \int B_\nu(T(r)) k_\nu E_2(\tau_\nu) d\nu,
\label{eq_cool_rate}
\end{equation}
where 
$N_{CO}$ is the local CO number density, 
$T$ is temperature, 
$k_{\nu}$ and $\tau_{\nu}$ are the absorption coefficients and zenithal opacity at frequency $\nu$, and 
$E_2(\tau)$ is the second exponential integral. 
The integral runs over the entire mm/submm range  (0-200 cm$^{-1}$), and 
unlike in \cite{str96}, We include all isotopic variants of CO and HCN, i.e. lines of CO, 13CO, C18O, HCN, H13CN and HC15N
are taken into account when calculating the opacities.
Moreover, absorption coefficients  are calculated using a Voigt profile, instead of the Doppler approximation. 
Both aspects lead to a minor but non-negligible increase in the cooling rate at low altitudes. 
Calculations of the cooling rates are performed for the thermal profile inferred in this work.

Results are shown in Fig.~\ref{fig_cool} for a series of assumed CO and HCN profiles.
The CO mixing ratio $q_{\rm CO}= 5 \times 10^{-4}$ curves show the ``nominal" CO cooling.
Although \cite{zal11a,zal11b} do not show their cooling rates, 
our calculation for CO can be compared to Figure 5a of \cite{zhu14}, showing reasonable agreement. 
Increasing the $q_{\rm CO}$ by a factor of 40 leads to an increase of the cooling rate, 
although in much lower proportion due to opacity effects. 

Other curves in Fig.~\ref{fig_cool} show 
the cooling due to HCN for different assumed HCN mixing ratios in the non-saturated part of the atmosphere. 
For the temperature profile derived in this work, the lower temperatures above 1270~km radius,  severely restrict 
the amount of gaseous HCN if saturation of HCN is accounted for. 
In fact, HCN appears to be saturated everywhere in the atmosphere, except possibly over the 1210-1270~km range,
where the condensation law allows $10^{-7}$-$10^{-6}$ HCN mixing ratios. 
There, the HCN cooling rate may slightly exceed the nominal CO cooling rate 
(pink vs. red curves in the left panel of Fig. \ref{fig_cool}). 
However, for the HCN cooling rates to approach the ``enhanced"  CO cooling rates necessary 
to explain our negative mesospheric temperature gradient 
(i.e. those for $q_{\rm CO} = 200 \times 10^{-4}$, as considered by  \cite{zal11a}),
one must assume that HCN is not limited by saturation.
Specifically, the blue curve in Fig.~\ref{fig_cool} shows that a uniform HCN mixing ratio of $\sim 5 \times 10^{-5}$ 
is required.

Although a full re-assessment of the radiative-models should be undertaken at this point, we conclude from this exercise that 
there is no obvious ``culprit" for a -0.2~K km$^{-1}$ temperature gradient above the radius $\sim$ 1220~km. 
According to the calculations of \cite{zal11a}, CO in amounts consistent with the direct observations
of \cite{lel11} provide unsufficient cooling. 
We show here that HCN could be an alternative efficient cooling agent, 
but only if its mixing ration vastly exceeds expectations from the condensation law. 
Direct measurements/upper limits of HCN from ALMA and perhaps from New Horizons
UV spectrometer (ALICE) will bring new light on this issue.\footnote{The detection of HCN in Pluto's atmosphere, using ALMA, was announced by \cite{lel15b} on July 30, 2015.}

\subsection{Alternative explanations}

Coming back a step, the primary result derived from a stellar occultation light-curve is the refractivity profile $\nu(r)$,
from which a density profile $n(r)= \nu(r)/K$ is obtained, assuming a given gas composition (Eq.~\ref{eq_nu_r}).

A first idea is to envisage that hazes are present in the mesosphere.
Those hazes would absorb part of the stellar flux, in such a way
that a basically isothermal mesosphere is thought to host a negative thermal gradient 
just because of the clear-atmosphere assumption. 
To test that hypothesis, we have generated synthetic light-curves, 
forcing the mesosphere to be isothermal at $T_{\rm iso}= 95.5$~K above the stratopause 
(we have also tested other values of $T_{\rm iso}$ between 85 and 110~K, with the same conclusions). 
Fig.~\ref{fig_fit_18jul12} shows the resulting residuals for the NACO 18 July 2012 light-curve
(labelled ``iso." in that figure).
They depart from zero well above the noise level, and
we can rule out photometric fluctuations caused by absorbing haze layers,
since the residuals have both positive and negative values.

This said, two assumptions may be wrong in Eq.~\ref{eq_n_r}: 
(1) the atmosphere may be not composed of pure nitrogen $N_2$, 
so that the nitrogen molecular mass $\mu$ must be replaced by a new value $\mu'$, and
(2) hypothetical zonal winds may create a centrifugal acceleration, so that 
the acceleration of gravity $g$ must be replaced by a term $g'$ that includes supplementary terms.

In fact, we can use  Eq.~\ref{eq_n_r} in a reversed way. 
More precisely, the refractivity profile $\nu(r)$ is actually an \it imposed \rm observable 
(since it is directly derived from the occultation light-curve), while we may use a 
\it prescribed \rm temperature profile $T'(r)$, for instance taken from a theoretical model.
With this approach in mind,  Eq.~\ref{eq_n_r} can be re-written:
\begin{equation}
\frac{\mu' g'}{\mu g} = -\frac{k T'}{\mu g} \frac{d\log (\nu T')}{dr}. 
\label{eq_mu_g}
\end{equation}
To obtain this equation,  we have used $\nu(r)= K \cdot n(r)$, 
where $K$ is the molecular refractivity (Eq.~\ref{eq_nu_r}). 
We assume here that the atmospheric composition varies slowly with radius 
(i.e. $K$ suffers small variations over one scale-height), 
so that we neglect $dK/dr$, and finally provide $d[\log(\nu)]/dr \sim d[\log(n)]/dr$.

Thus, the ratio $\mu' g'/\mu g$ is the factor by which the molecular mass and/or the acceleration of gravity $g$
must be multiplied in order to retrieve a prescribed temperature profile $T'(r)$, 
given an observed occultation light-curve.

In Fig.~\ref{fig_mu_g_r}, we consider an example where the prescribed temperature profile $T'(r)$ 
exhibits a decrease of only 10~K between the stratopause and the mesopause.
This is typical of what can be obtained by the combined effects of 
CO  cooling \citep{zal11a} and/or 
a atmospheric escape \citep{zhu14}.
The right panel of  Fig.~\ref{fig_mu_g_r} shows the resulting profile for  $\mu' g'/\mu g$, 
restricting ourselves to the mesospheric region.

We first assume here that the atmosphere is composed of pure nitrogen, so that $\mu'= \mu$, and
the ratio $\mu' g'/\mu g= g'/g$ is only caused by variations of $g'$. 
In the presence of a zonal wind with velocity $v$, the centrifugal acceleration provides 
$g'= g - v^2/r$, and from $g=GM/r^2$, a zonal wind of:
\begin{equation}
v= \sqrt{\frac{GM}{r}} \cdot \sqrt{1 - \frac{g'}{g}} \sim 840  \sqrt{1 - \frac{g'}{g}} {\rm~m~s^{-1}},
\label{eq_zonal}
\end{equation}
where we have used the value of $GM$ in Table~\ref{tab_param} and $r \sim 1,250$~km.
With the example above, the factor $g'/g'$ reaches a minimum value of about 0.95, 
yielding $v \sim 190$~m~s$^{-1}$.
This is close to supersonic, 
as the speed of sound for nitrogen $N_2$ at 100~K is about 200~m~s$^{-1}$.
In fact, current General Circulation Models (GCM's) for Pluto predict zonal winds of  less than
10~m~s$^{-1}$ at the altitudes considered here \citep{van13,zal13}.
Moreover, we see that above $r \sim 1,300$~km, the ratio $g'/g$ becomes larger than
unity with the example considered here, which is impossible from Eq.~\ref{eq_zonal}.
Other prescribed profiles $T'(r)$ could be imagined to avoid this problem by displacing 
the $\mu' g'/\mu g$ profile to the left in Fig.~\ref{fig_mu_g_r} (providing smaller values of $g'/g$), 
but this would imply even more unrealistic, high zonal winds.

Considering that $g'/g \sim 1$ from the discussion above, the  $\mu' g'/\mu g$ profile would 
represent variations of molecular of the atmospheric molecular weight, $\mu'/\mu$. 
In the example of Fig.~\ref{fig_mu_g_r}, the molecular weight of the atmosphere has to be
inferior to that of molecular nitrogen, $\mu$, to mimic the effect of a negative temperature gradient.
This could be caused by the presence of a lighter gas, for instance neon, which has a
molecular weight $\mu_{\rm Ne} \sim 0.72 \mu$.
That species has a relatively large solar abundance ($N_e/N \sim 1.5$) and is not
condensed at Pluto's atmospheres temperatures.
The minimum value $\mu'/\mu = 0.95$ near 1,230~km  (Fig.~\ref{fig_mu_g_r}) 
would then require a local neon abundance of about 82\%.
However, and as before, the ratio $\mu'/\mu$ would be larger than unity above 1,300~km,
requiring that another, heavier, gas (e.g. argon) takes over above 1,300 km and drives 
the molecular mass upwards.
Such model is clearly unrealistic though, because mass separation would result in 
a depletion, not enrichment, of the heavier species in the upper atmosphere.

\section{Upper atmosphere}

Above $r \sim 1,400$~km, the results of Fig.~\ref{fig_T_r_invers_par12} shows that a change
of thermal gradient may occur, with a more isothermal upper branch just above the mesosphere.
However, the lack of independent constraints on the temperature at that level prevents an 
unambiguous choice for a particular solution for $T(r)$. 
In addition, the rapidly increasing contribution of the noise makes impossible any estimation of the 
thermal gradient above 1,450~km (Fig.~\ref{fig_T_r_invers_par12}, panel (d)).
This said, the lack of obvious mechanisms to drastically warm up or cool down the atmosphere just above 1,400~km
suggests (but by no means proves) an isothermal branch between 1,400 and 1,450~km.
Under this hypothesis, we estimate a 1-$\sigma$ uncertainty domain of $81\pm 6$~K for the 
temperature of this isothermal branch.  
(This interval corresponds to an increase of
$\Delta \chi^2 = +1$ of the $\chi^2$ function  with respect to the best, minimum value $\chi^2_{\rm min}$).

\section{Discussion and conclusions}

\begin{figure*}[!htb]
\centering  
\includegraphics[height=6cm,trim=50 0 -20 30]{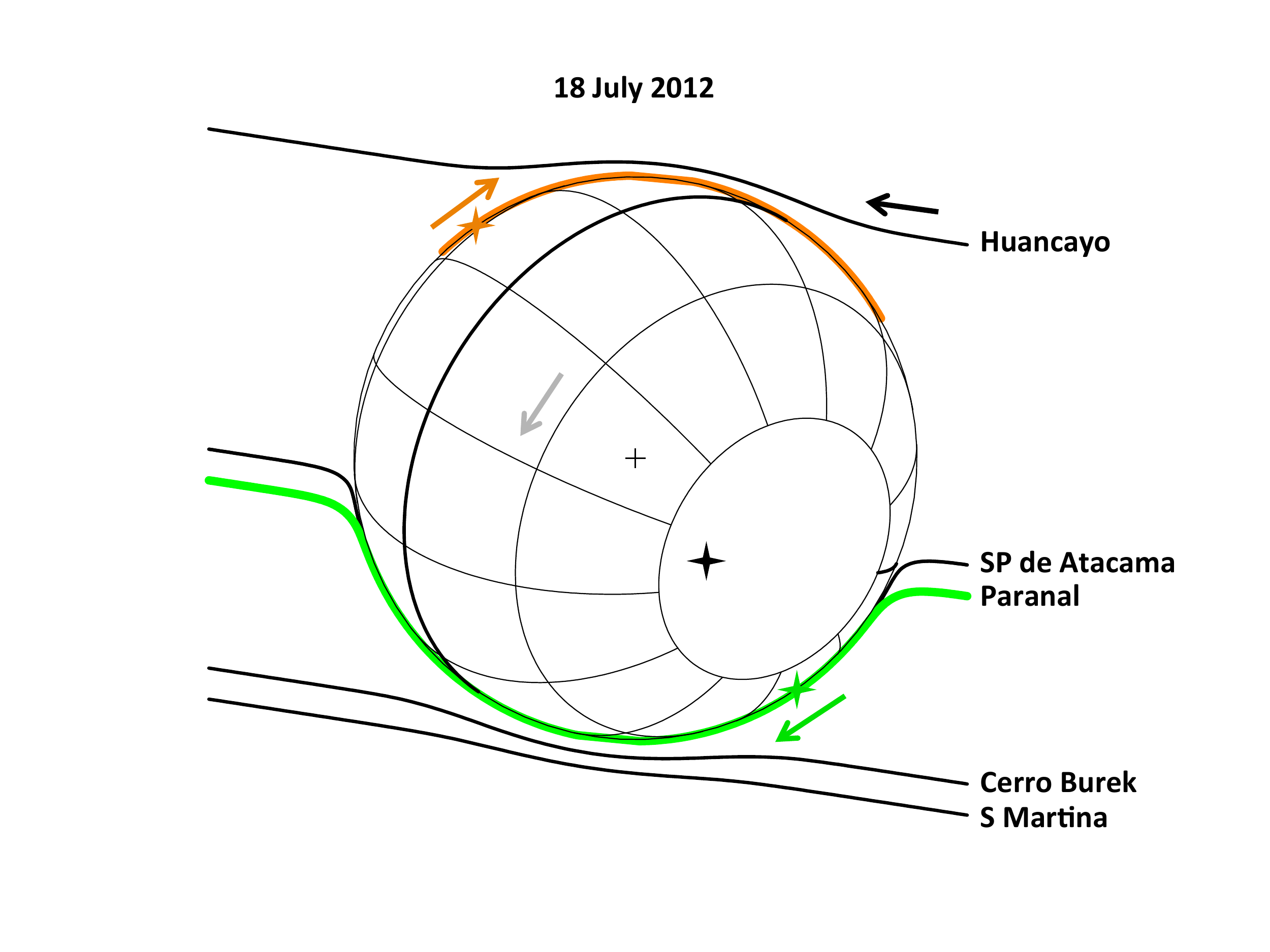}
\includegraphics[height=6cm,trim=0   0  20 30]{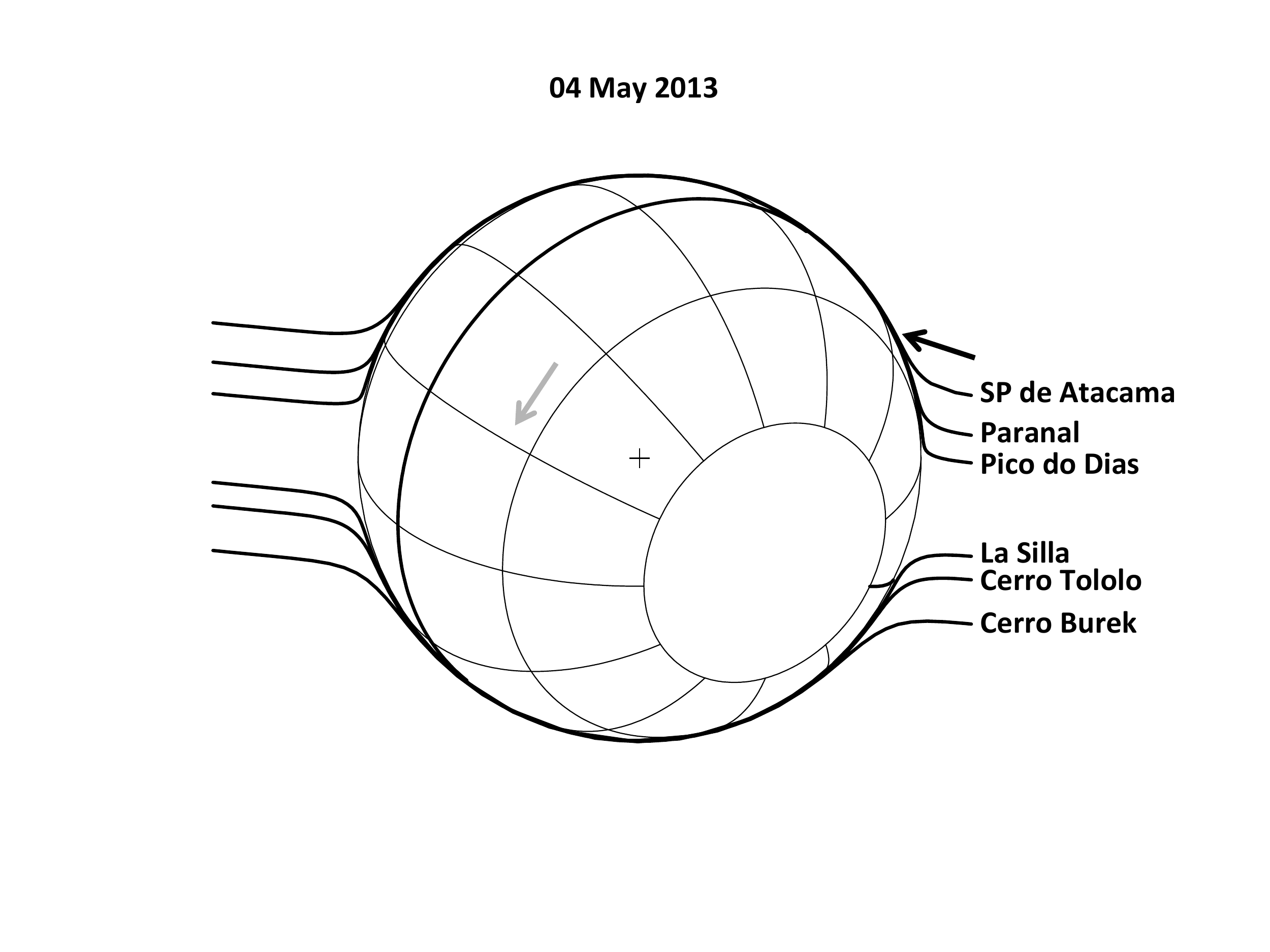}
\caption{%
\normalsize
Left -
The trajectories of the primary stellar images relative to Pluto, 
as seen from the five stations used on 18 July 2012, see Table~\ref{tab_18jul}. 
The black arrow shows the general direction of stellar motion.
Here, Pluto's has an assumed radius $R_P=1,190$~km (see text), and its 
center is indicated by the cross symbol.
The gray arrow inside the disk indicates the direction of rotation. 
In the case of Paranal, we have plotted the path of the primary image in green,
\textit{and} 
the associated path of the secondary image in orange (see also Fig.~\ref{fig_fit_18jul12_NACO}). 
The green and orange arrows show the corresponding local stellar motion along Pluto's limb. 
Note that the two images move in opposite directions.
The black star symbol shows the star position as seen from Paranal at a given, arbitrary moment, while 
the green and orange star symbols indicate the associated primary and secondary images at
that time, respectively.
Note that the three star symbols and the cross are aligned.
Right~-
The same as left panel for the 04 May 2013 occultation, 
with only the paths of the primary stellar images plotted. In both panels, the summer, permanently lit Pluto's hemisphere is at right, and the low insolation winter limb is at left.
}
\label{fig_XY}
\end{figure*}

\begin{figure*}[!htb]
\centering
\includegraphics[height=16.5cm]{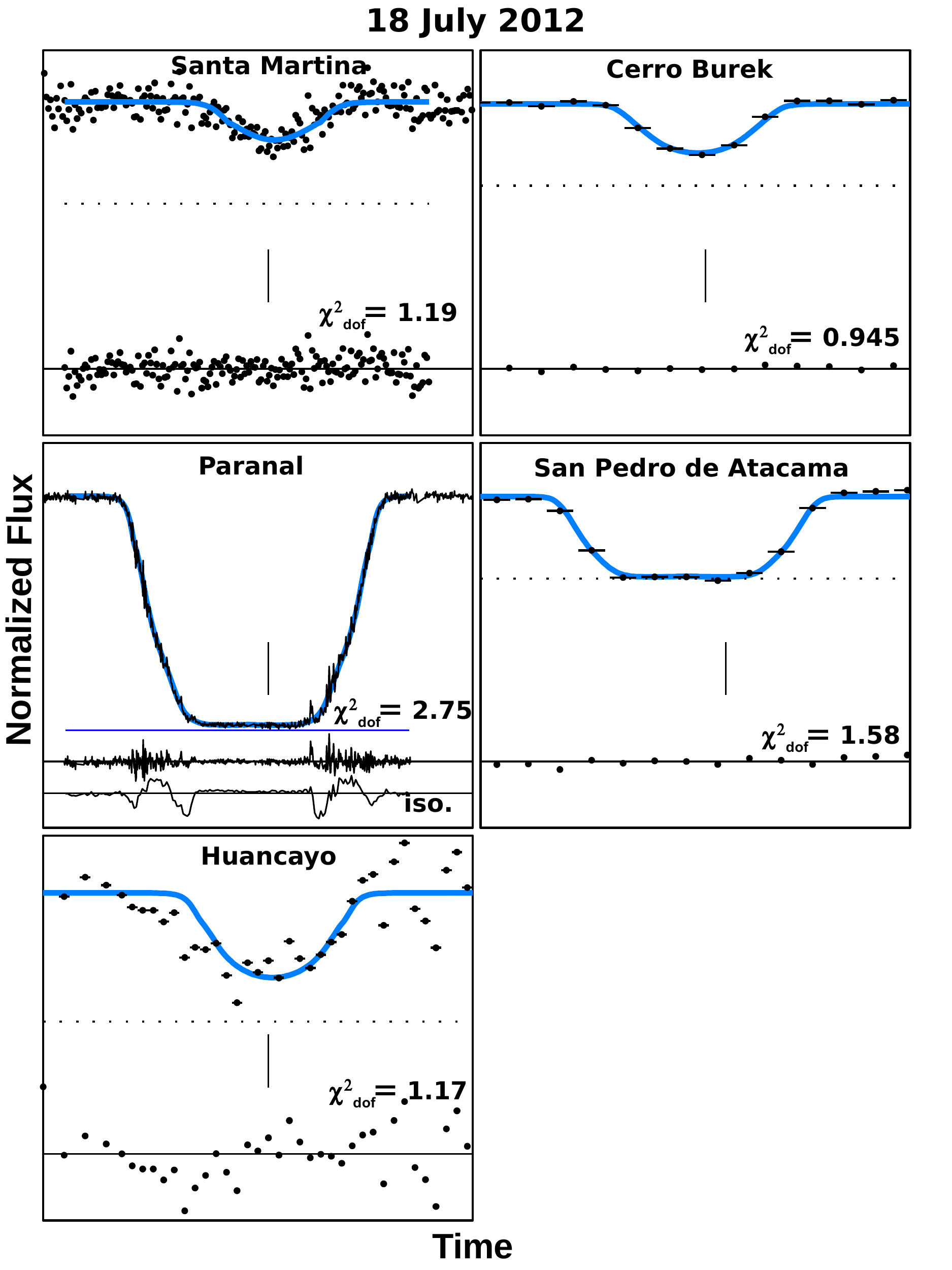}
\caption{%
The blue curves are a simultaneous fit to the 18 July 2012 light-curves,
using the best atmospheric model described in Table~\ref{tab_atmo_fit} and Fig.~\ref{fig_model}.
The number at the lower right of each panel is the value of $\chi^2_{\rm dof}$ (Eq.~\ref{eq_chi2}), 
i.e. the $\chi^2$ per degree of freedom for each corresponding fits. 
Each panel spans 3 minutes of data, with the vertical tick marks located at 04:13 UT.
All the light curves show the total flux (star+Pluto+Charon) plotted at the same vertical scale.
The horizontal bars on the Cerro Burek, San Pedro de Atacama and Huancayo data points 
represent the respective integration times.
The zero flux is indicated by the solid horizontal line at the bottom of each panel,
together with the residuals (data minus model).
The dotted horizontal lines mark the \it fitted \rm zero stellar fluxes 
(or equivalently, the Pluto+Charon contribution to the total flux), 
obtained using our best Pluto atmospheric model.
The blue horizontal line in the Paranal panel marks the \it measured \rm zero stellar flux at that station, 
the only one at which a photometric calibration was possible 
(see text and Figs.~\ref{fig_coro}, \ref{fig_fit_18jul12_NACO}).
In the Paranal panel, we have also added the residuals (labelled ``iso.") obtained by forcing
an isothermal mesosphere at $T_{\rm iso}= 95.5$~K. The residuals have been averaged over 5-s time 
intervals and shifted vertically by -0.12 for better showing the clear discrepancy between 
the isothermal mesospheric model and the data.
Other values chosen for  $T_{\rm iso}$ would result in the same qualitative behavior.
In essence, isothermal mesospheres do not provide satisfactory fits to the NACO light-curve.
}%
   \label{fig_fit_18jul12}
\end{figure*}

\begin{figure*}[!htb]
\centering
\includegraphics[height=17.5cm]{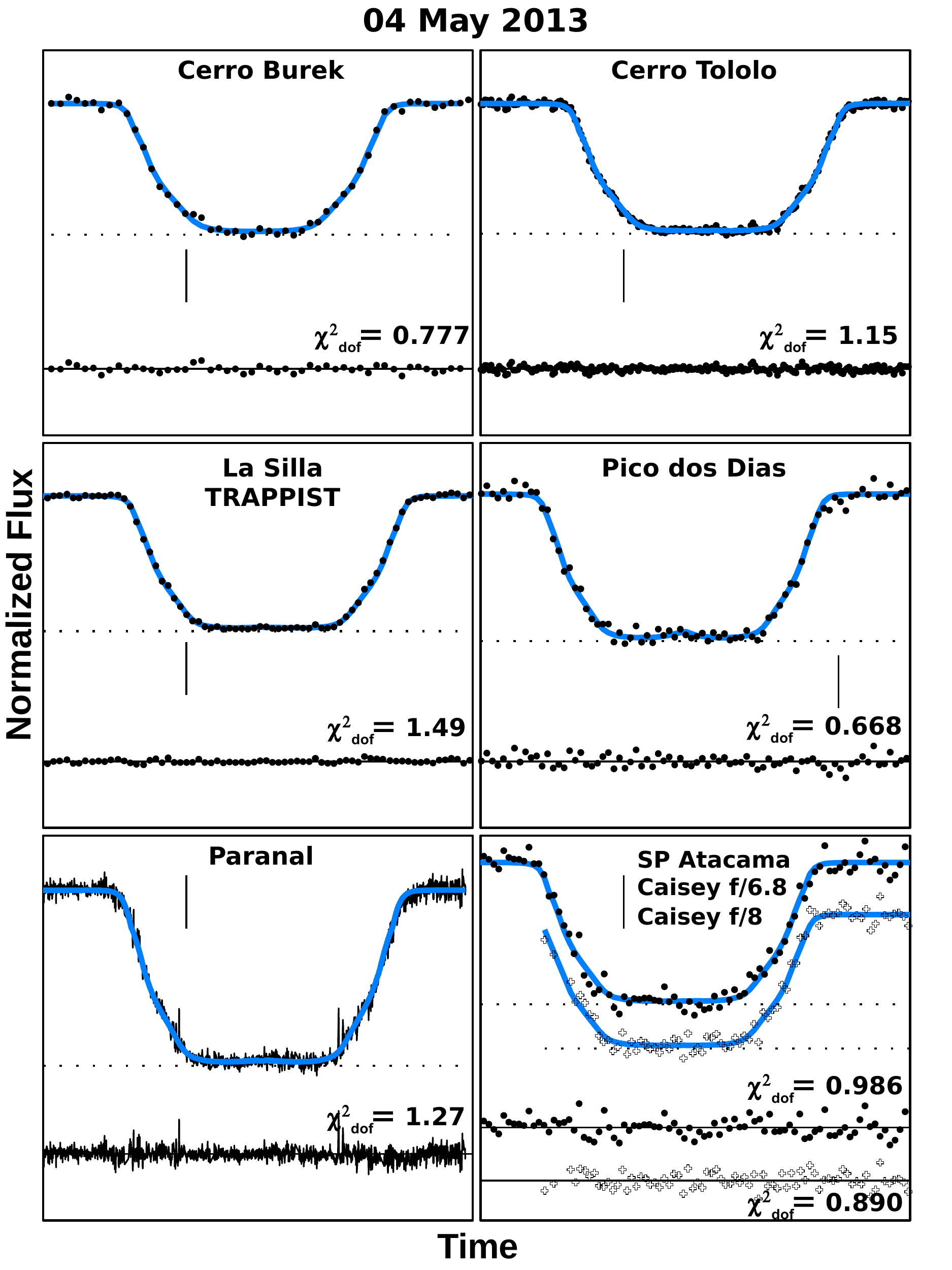}
\caption{%
\normalsize
The same as Fig.~\ref{fig_fit_18jul12} for the 04 May 2013 event.
Each panel now represents 6 minutes of data, with  
the vertical tick mark located at 08:22~UT.
Note that the two light-curves from San Pedro ( ``SP") de Atacama have been 
displaced vertically by $\pm 0.1$ for better viewing.
}%
   \label{fig_fit_04may13}
\end{figure*}

\begin{figure*}[!htb]
\centering 
\includegraphics[height=8cm]{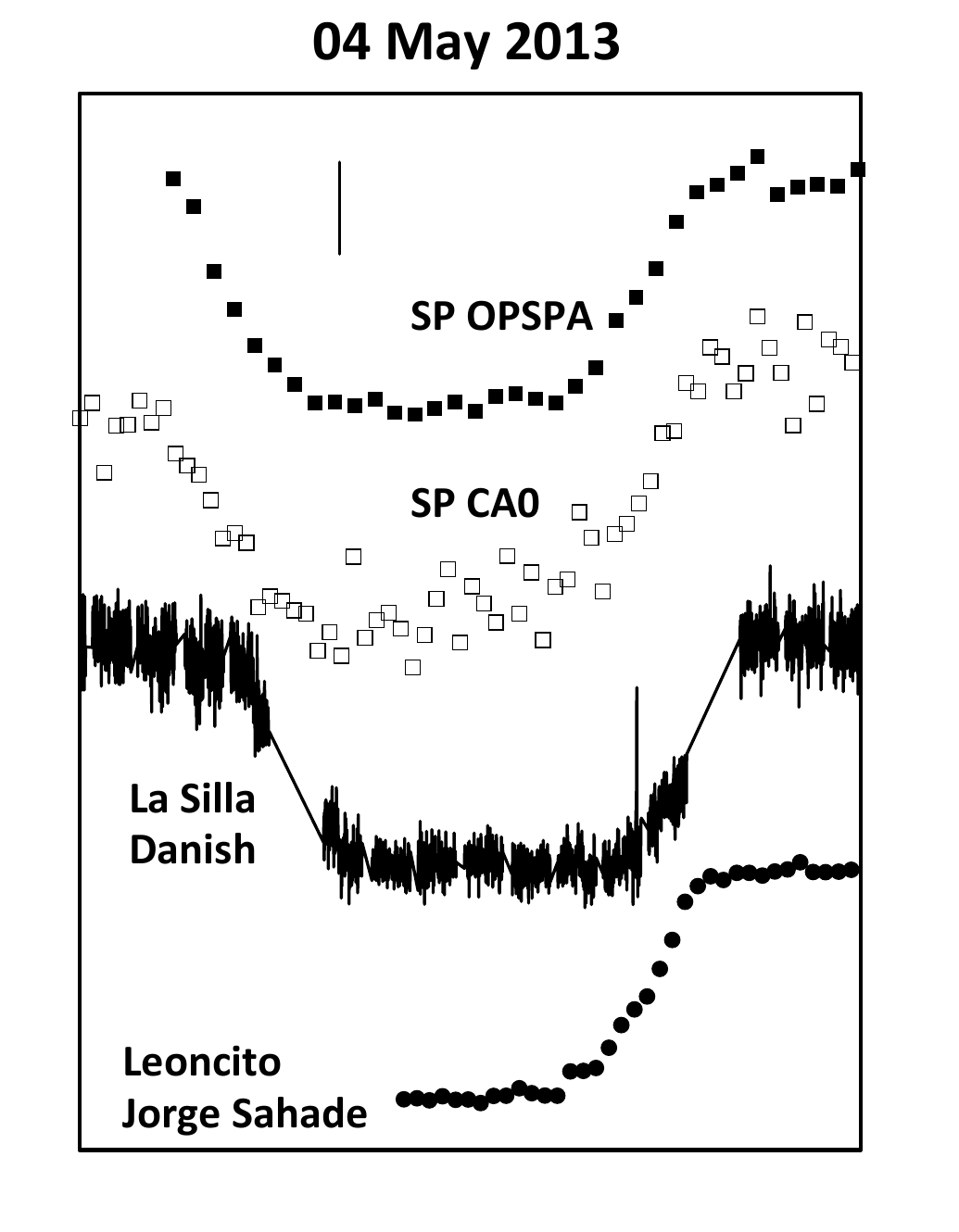}
\caption{%
\normalsize
The same as Fig.~\ref{fig_fit_04may13}, but for the light-curves that were not
included in the fit, either due to lower SNR, or interruptions during the acquisition.
See Table~\ref{tab_04may} for intrumental details 
(``SP" refers to San Pedro de Atacama and acronyms refer to telescope used in that station.).
Note that the Leoncito, Danish and SP light-curves duplicate the observations of the
Cerro Burek, 
La Silla TRAPPIST and
Caisey telescopes, respectively.
}%
\label{fig_leo_dan_spa}
\end{figure*}

\begin{figure*}[!htb]
\centering
\includegraphics[width=7cm, height=8cm]{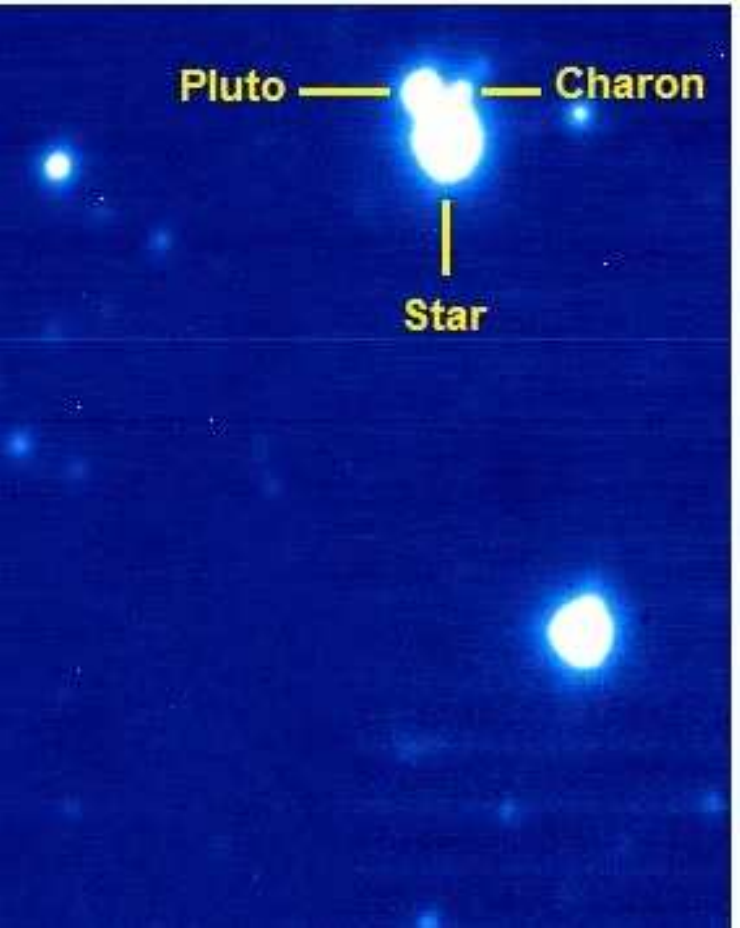}
\includegraphics[width=7cm, height=8cm]{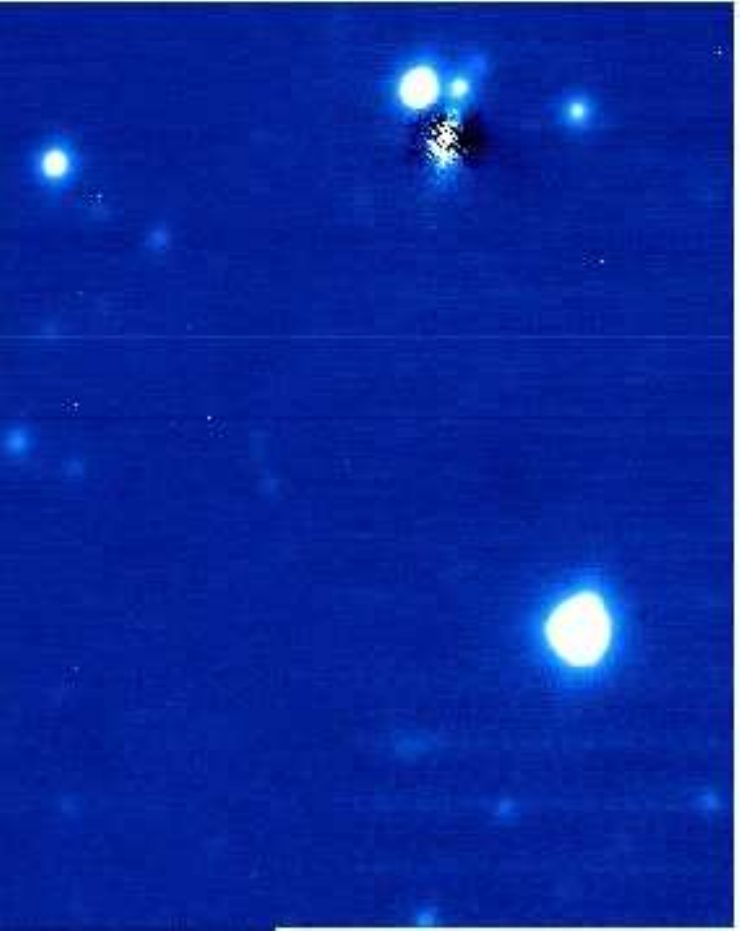}
\caption{%
\normalsize
Photometric calibration of the July 18, 2012 event (Paranal/VLT, NACO H-band). 
Left  - 
Image taken some 20 minutes before the event, showing the small separation between Pluto, Charon and the star ($\sim$ 1'').
Right - 
The same image after a digital coronagraphy treatment that removed the stellar image.
See text for details.
}%
\label{fig_coro}
\end{figure*}

We have analyzed among the best light-curves ever obtained during stellar occultations by Pluto.
Combination of well-sampled occultation chords (Fig.~\ref{fig_XY}) and
high SNR data (Figs.~\ref{fig_fit_18jul12},\ref{fig_fit_04may13}) have allowed us to constrain the  
density, temperature and thermal gradient profiles of Pluto's atmosphere between radii 
$r \sim 1,190$~km (pressure $p \sim 11$~$\mu$bar) and
$r \sim 1,450$~km (pressure $p \sim 0.1$~$\mu$bar).
Our main results are listed below.

\begin{figure*}[!htb]
\centering
\includegraphics[height=11cm,trim=100 0 0 0]{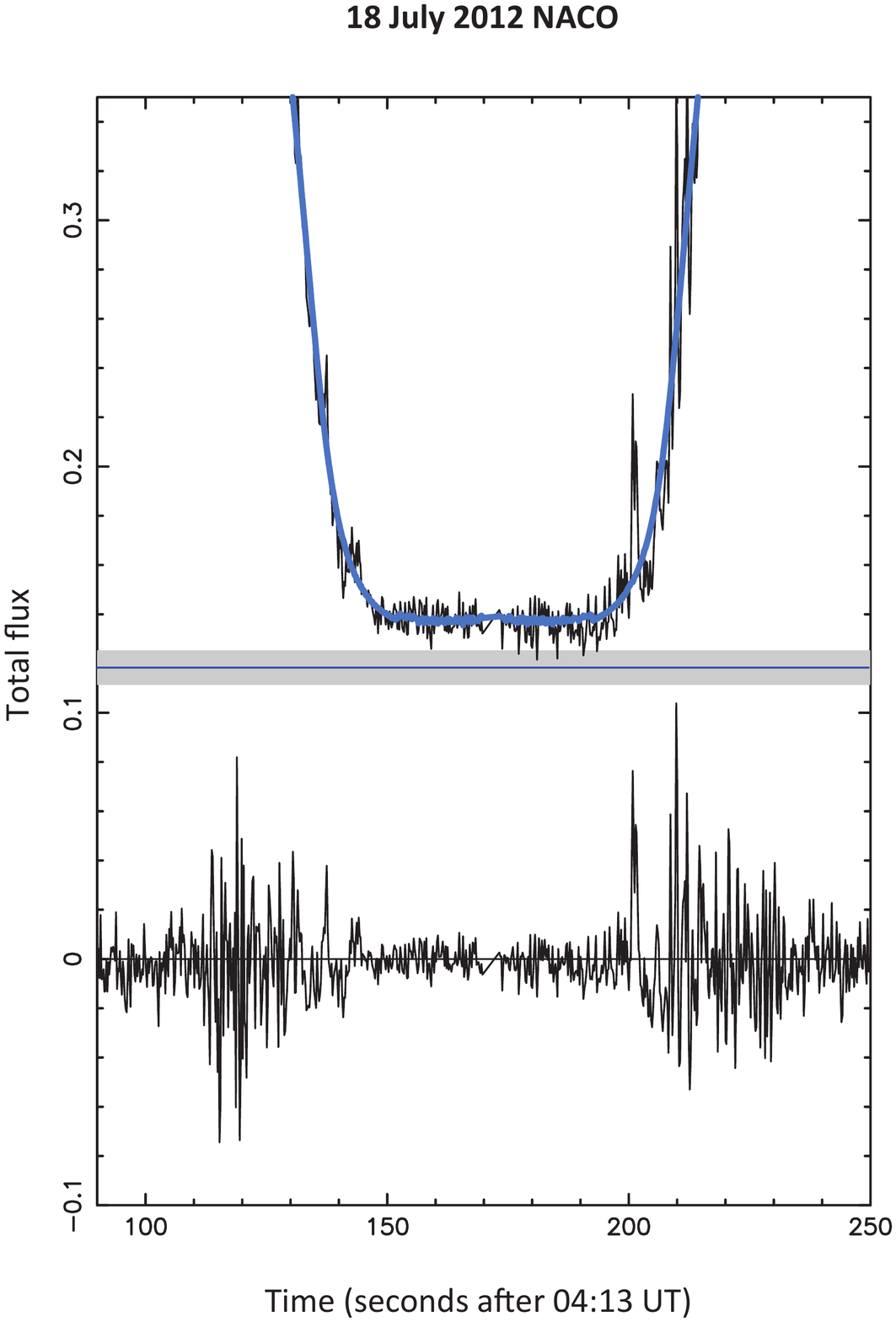}
\includegraphics[height=11cm,trim=0 0 100 0]{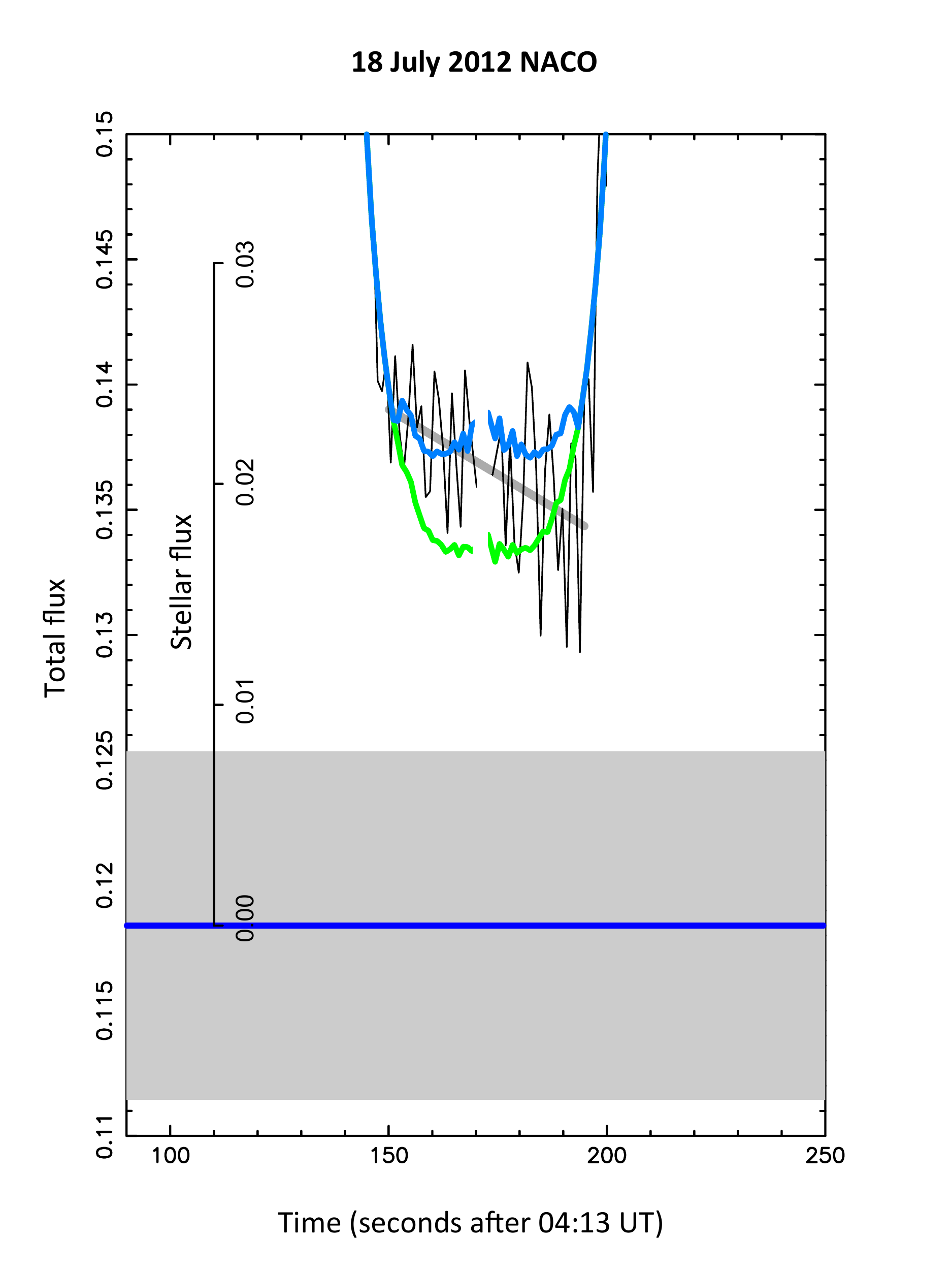}
\caption{%
\normalsize
Left - 
details of the fit to the NACO 18 July 2012 light-curve 
(see also the middle-left panel of Fig.~\ref{fig_fit_18jul12}).
The horizontal blue line in the gray shaded area indicates the Pluto + Charon contribution
to the total observed flux and its 1-$\sigma$ error bar,  $0.1184 \pm 0.007$.
The residuals curve at the bottom clearly shows the spike activity at ingress and egress. 
Right - 
Expanded view of the left panel. 
The data have been binned over 1 second-time intervals to better show the flux decrease 
during the central phase of the occultation. 
The flux of the \textit{primary} stellar image is plotted in green, while the blue curve
is the sum of the primary and secondary images, according to the model 
(see Fig.~\ref{fig_XY} and Appendix). 
Thus, the contribution of the secondary image is the difference between the blue and green curves.
Note the interruption of data acquisition (about 3 s) at mid-occultation,
necessary to the writing of the data cube before the start of the next data cube.
The inclined gray line is a linear fit to the central part of the light-curve,
which illustrates the ingress/egress asymmetry of the residual stellar flux.
The vertical axis inside the box at left indicates the value of the residual, normalized stellar flux.
It shows that the stellar flux decreased from about 2.3\% to about 1.8\% of its full
unocculted value during that interval. 
The systematic error on those values is $\pm 0.8$\% (corresponding to the shaded area).
}%
   \label{fig_fit_18jul12_NACO}
\end{figure*}

\begin{figure*}[!htb]
\centering
\includegraphics[height=8.3cm]{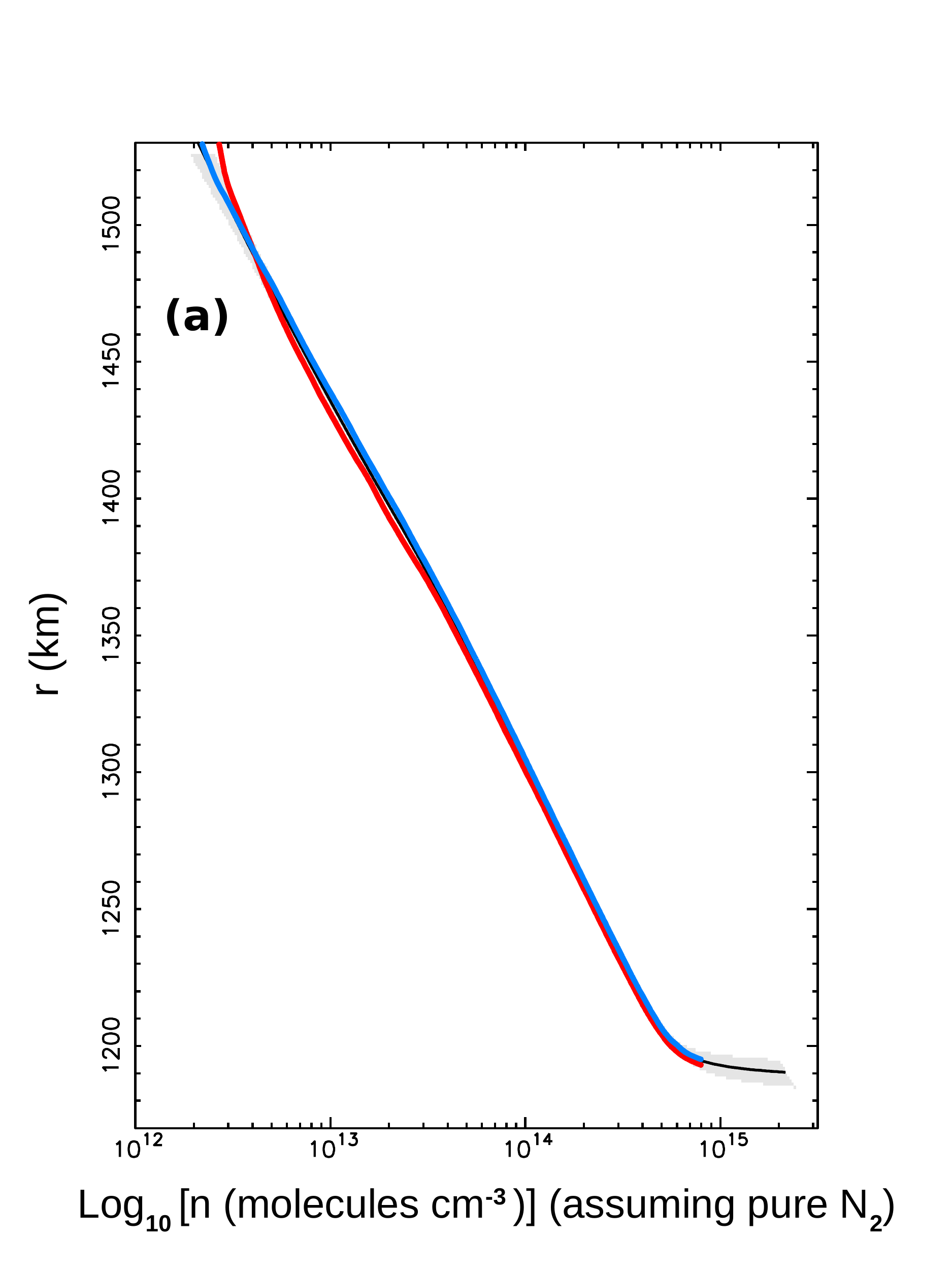}
\includegraphics[height=8.3cm]{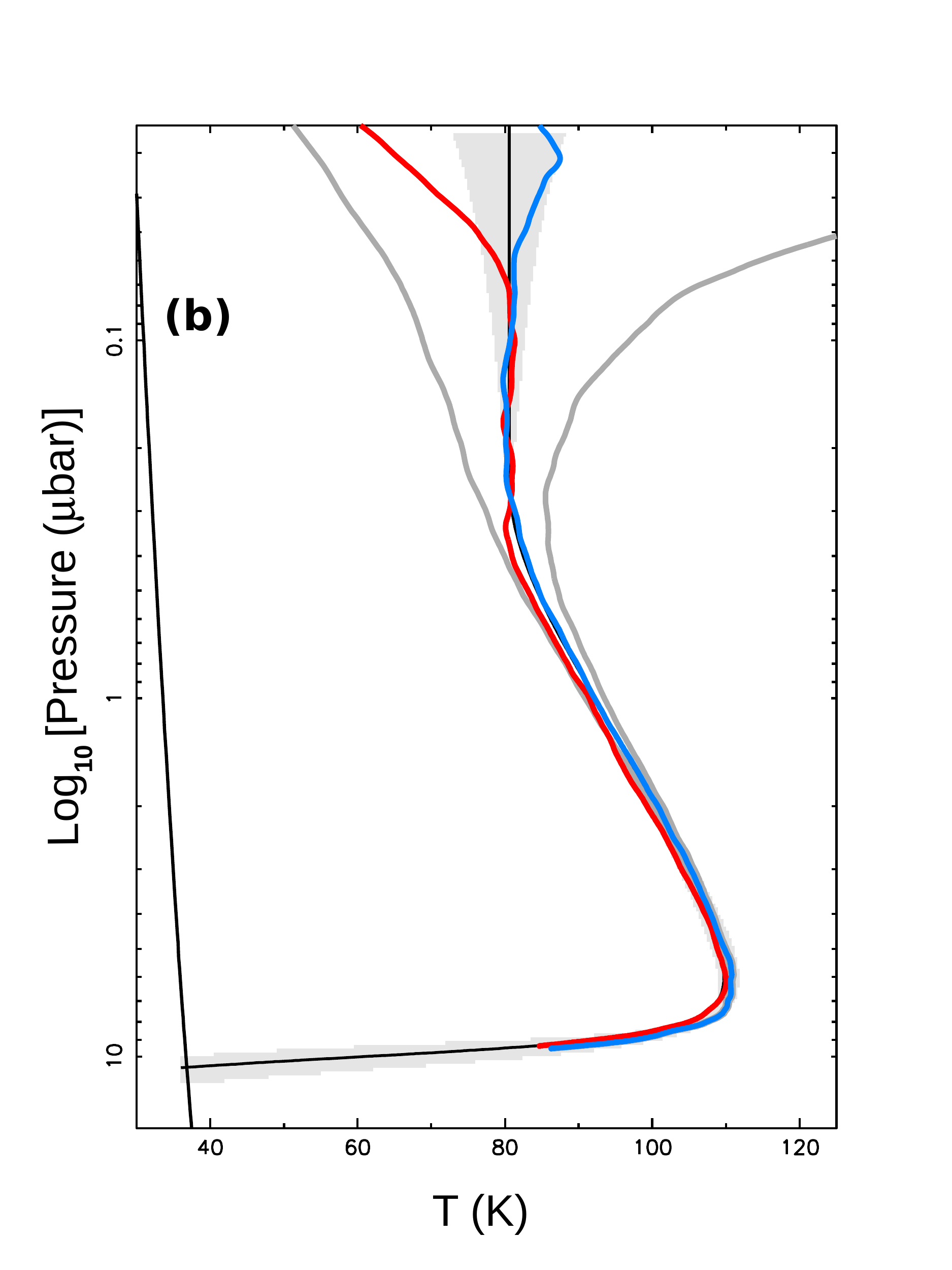}
\includegraphics[height=8.3cm]{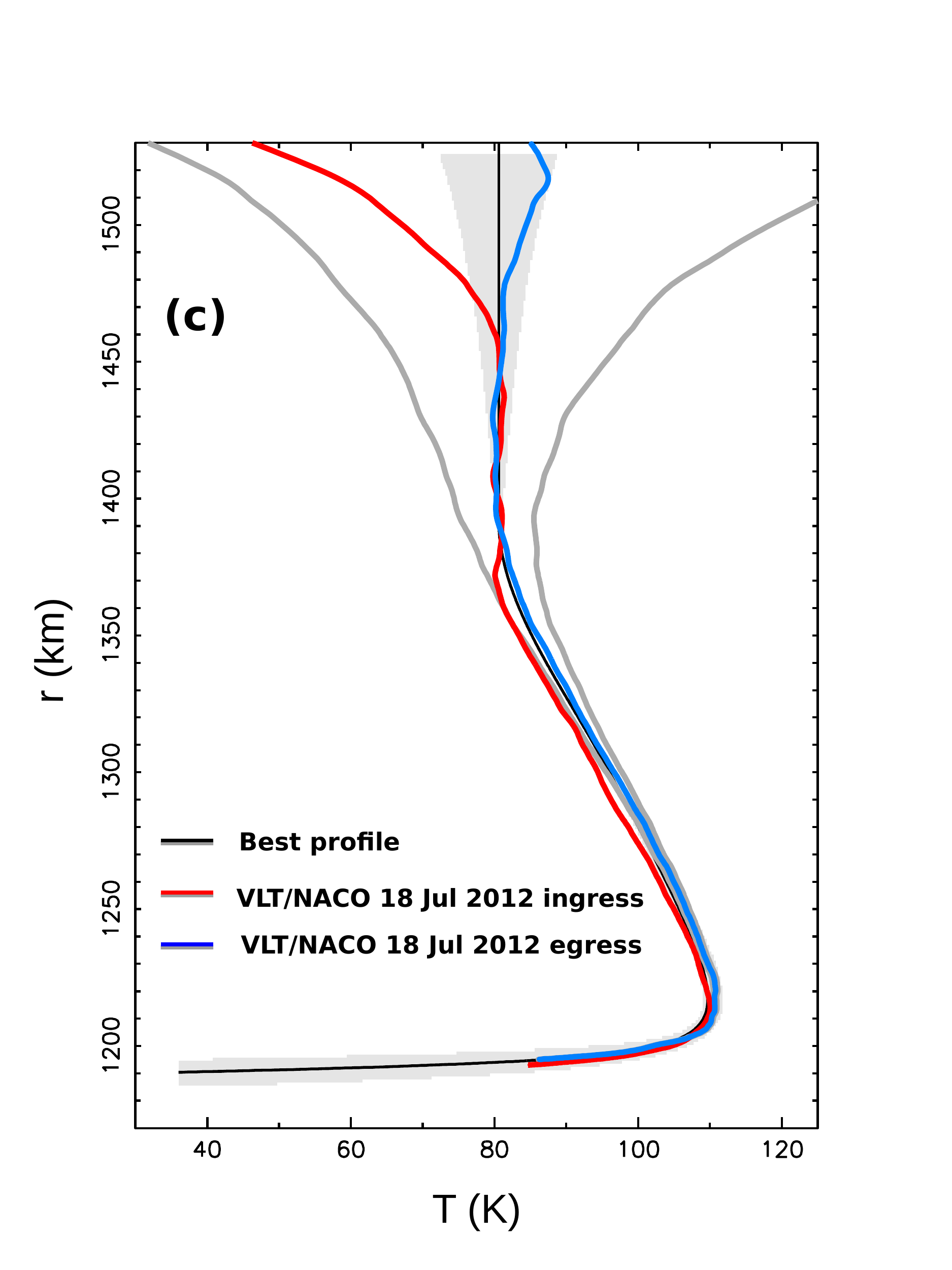}
\includegraphics[height=8.3cm]{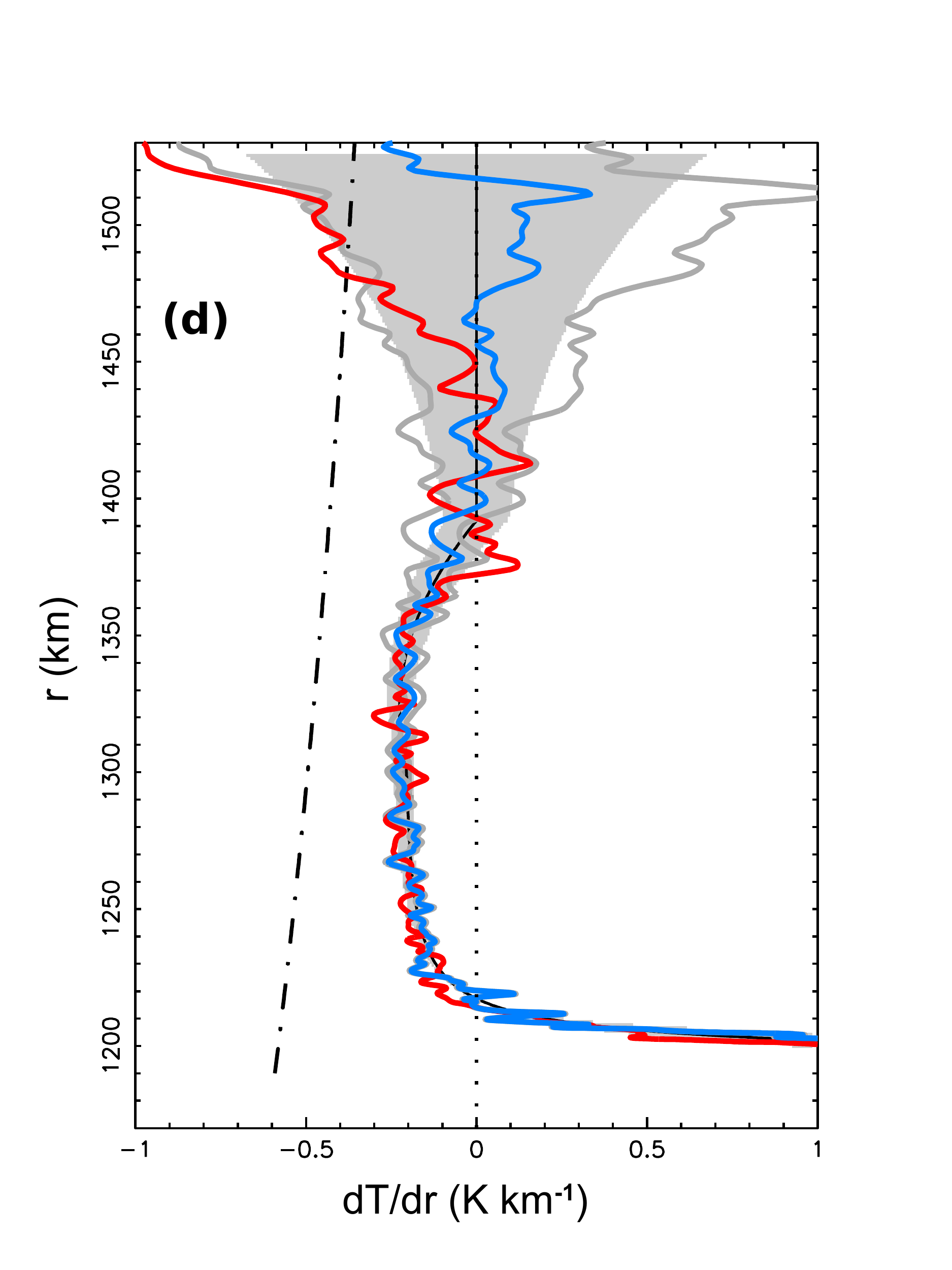}
\caption{%
\normalsize
In all the panels,
the black solid line is the model that best fits all the 18 July 2012 NACO light-curves, 
see Figs.~\ref{fig_fit_18jul12} and Table~\ref{tab_atmo_fit}.
The red (resp. blue) lines are the particular profiles obtained from the inversion of 
the NACO 18 July 2012 light-curve at ingress/summer (resp. egress/winter). 
The shaded areas at the top of the profiles indicate the expected $\pm$1$\sigma$
fluctuations caused by the photometric noise, see text.
The shaded areas at the bottom of the profiles are the $\pm$1$\sigma$ uncertainty
domain caused by the uncertainty on the Pluto + Charon contribution 
to the 18 July 2012  NACO light-curve, see Fig.~\ref{fig_fit_18jul12_NACO}.
(a) Molecular density vs. radius (assuming a pure $N_2$ atmosphere); 
(b) temperature vs. pressure; 
(c) temperature vs. radius; 
(d) temperature gradient vs. radius.
The two gray temperature profiles in panels (b), (c) and (d) show the effect of different temperature 
boundary conditions for the egress NACO profile.  
More precisely, those profiles differ from the nominal one (blue lines) by $\pm 5$~K at 1,390~km.
The oblique solid line at the left of panel (b) is the vapor pressure equilibrium limit for $N_2$ \citep{fra09}.
Nitrogen should condense at the left of this line, so that its intersection with the temperature profile
may define Pluto's surface in the absence of troposphere, 
see Fig.~\ref{fig_T_r_invers_par12_bottom} and text for details. 
The dash-dotted line in panel (d) is the dry adiabat for a pure $N_2$ atmosphere.
}%
\label{fig_T_r_invers_par12}
\end{figure*}

\begin{figure*}[!htb]
\centering
\includegraphics[height=9cm]{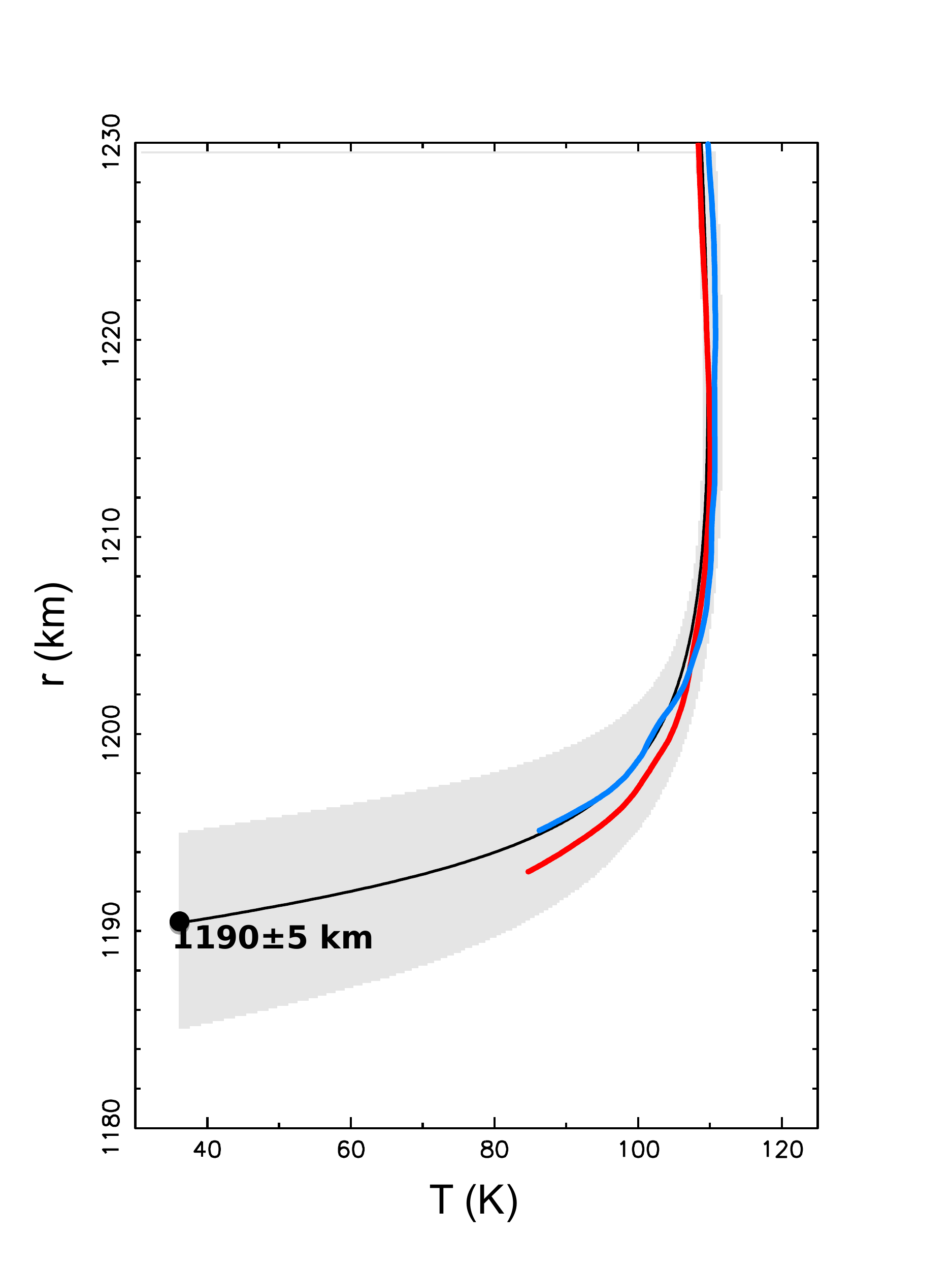}
\includegraphics[height=9cm]{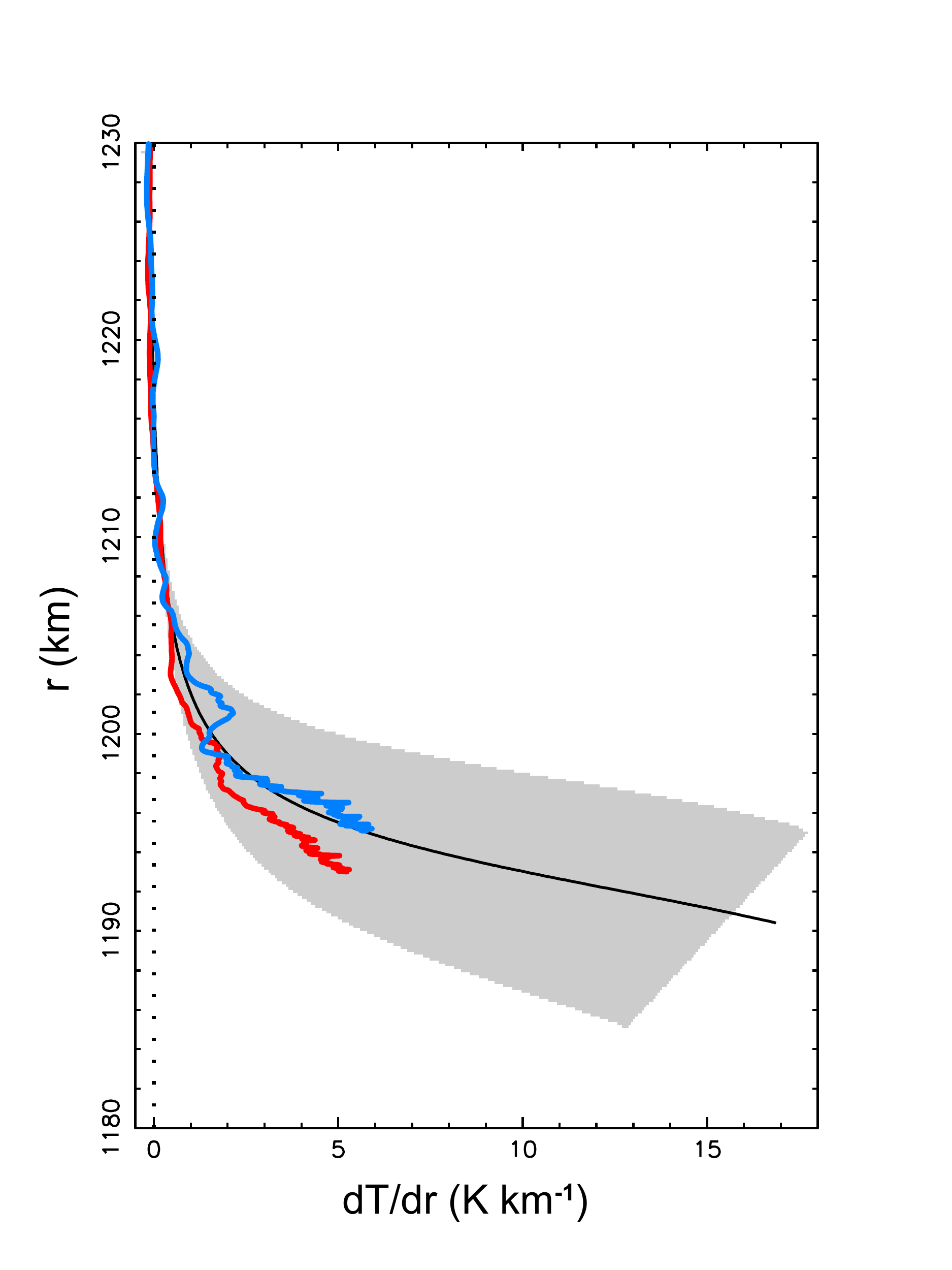}
\caption{%
\normalsize
Left - 
Expanded view of the bottom of the temperature profiles shown in Fig.~\ref{fig_T_r_invers_par12}.
The bullet is the intersection with the nitrogen condensation line. 
The error bar attached to its positions is defined by the radial extension 
of the shaded uncertainty domain.
Right -
The corresponding expanded view for the temperature gradient.
}%
\label{fig_T_r_invers_par12_bottom}
\end{figure*}

\begin{figure*}[!htb]
\centering
\includegraphics[width=12cm]{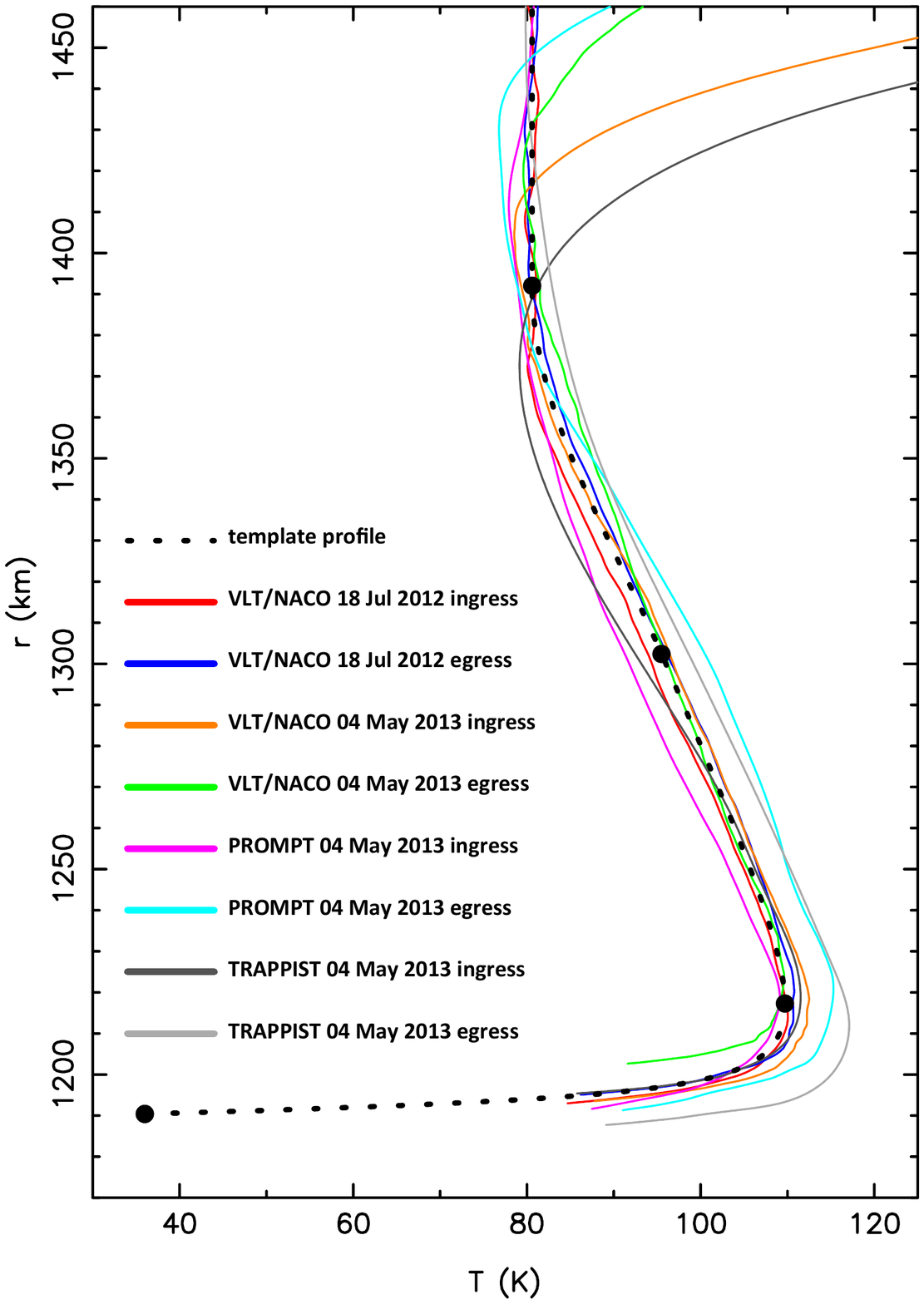}
\caption{%
\normalsize
The temperature profiles derived from the inversion of our best occultation light-curves obtained on 
18 July 2012 and 04 May 2013. 
The dotted line is our global, best-fitting temperature profile 
(also shown in Figs.~\ref{fig_T_r_invers_par12} and \ref{fig_model}).
}%
\label{fig_T_r_invers}
\end{figure*}

\begin{figure*}[!htb]
\centering 
\includegraphics[width=17.0cm, height=7.0cm]{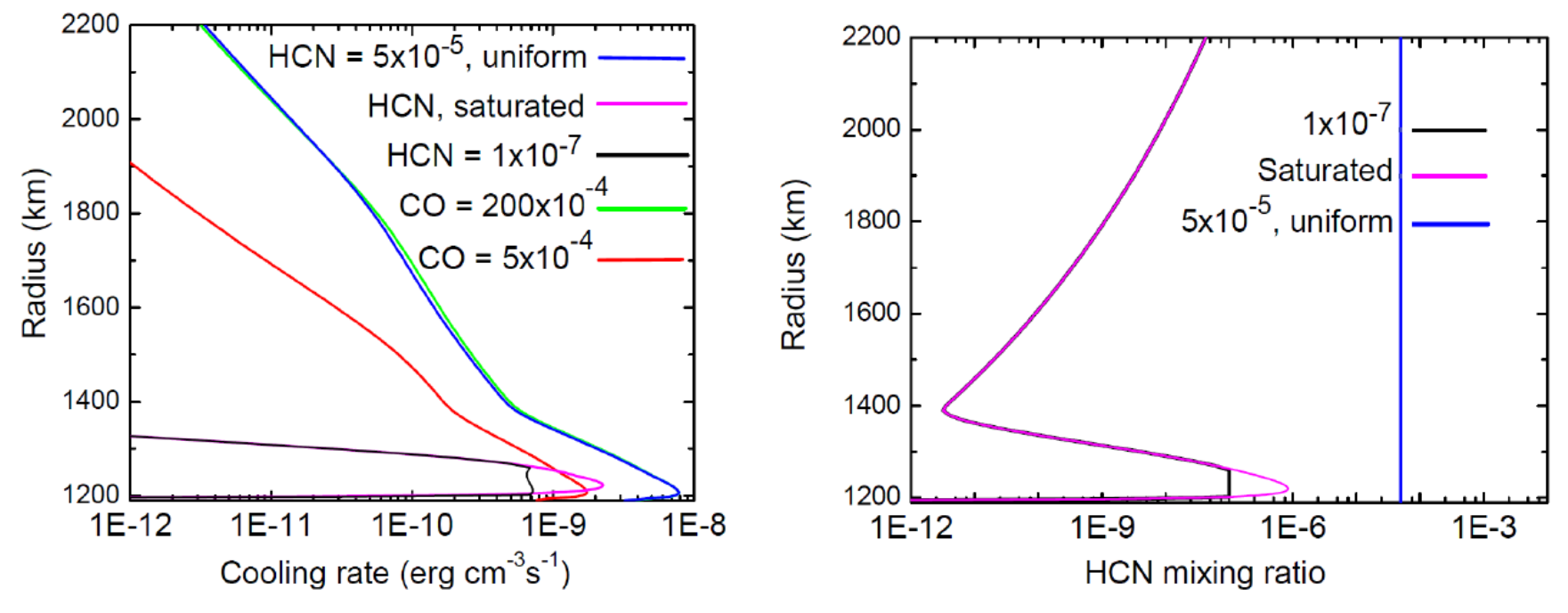}
\caption{%
\normalsize
%
Calculation of cooling rates by CO and HCN. 
The y-axis is the distance from Pluto's center, with the surface position assumed here at 1,184~km.
Left panel - Cooling rates assuming the thermal profile from this work. 
Red and green curves: CO cooling rates for $q_{\rm CO}= 5 \times 10^{-4}$ and $200 \times 10^{-4}$, respectively.
The other three colored curves show the HCN cooling rate for the corresponding HCN profiles.
Right panel -  HCN mixing ratios profiles. The black and purple curves make use of the thermal profile from this work. 
Due to the significantly cold temperatures above $\sim$1300~km, HCN is limited by saturation throughout the atmosphere, 
except in a limited region at 1210-1270~km for an assumed $q_{\rm HCN}= 10^{-7}$. 
The blue curve shows the hypothetical case of a uniform (i.e. non-limited by saturation) 
$5 \times 10^{-5}$ uniform HCN mixing ratio. 
%
%
%
%
%
%
%
}%
\label{fig_cool}
\end{figure*}

\begin{figure*}[!htb]
\centering 
\includegraphics[height=10cm]{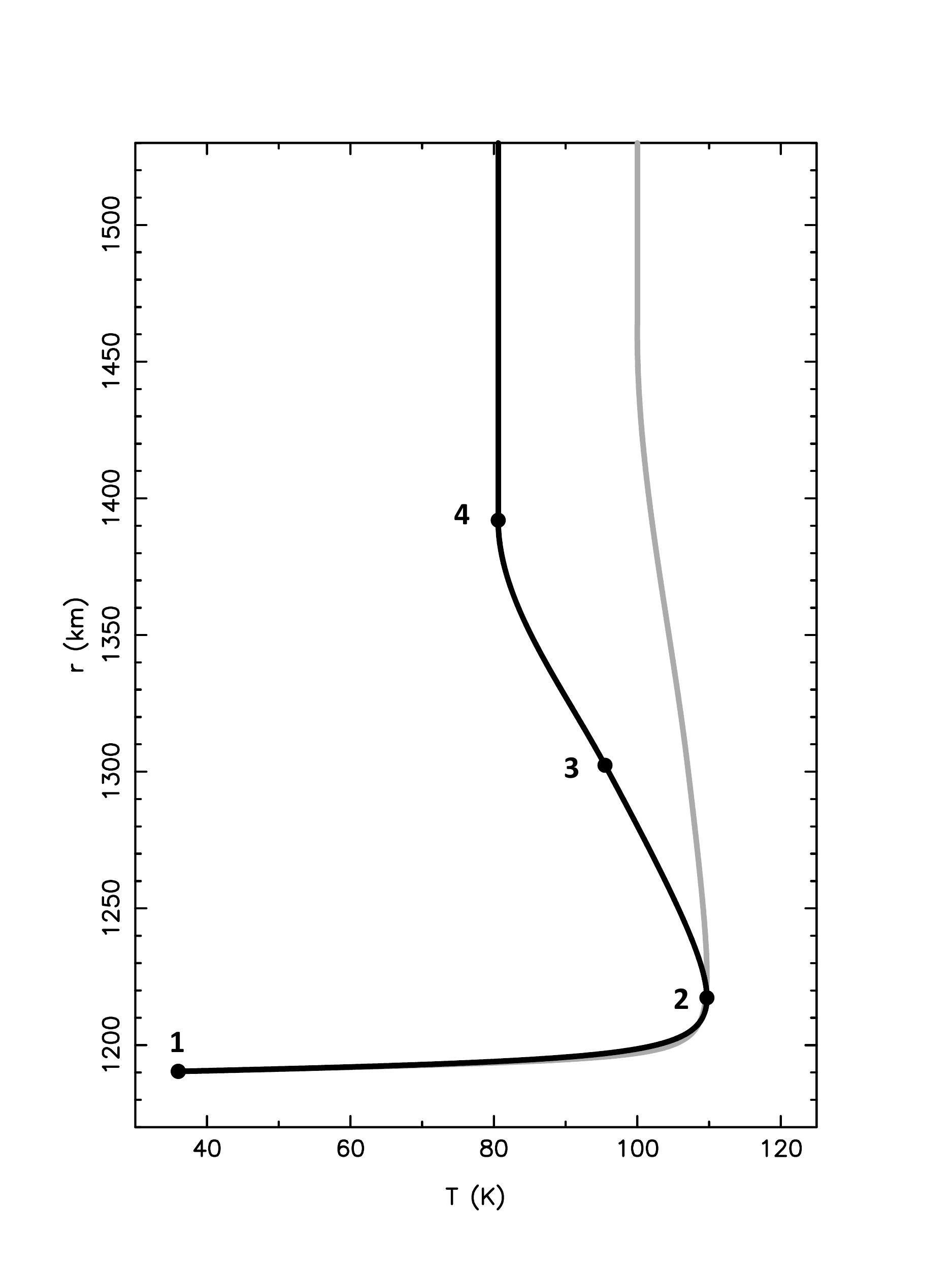}
\includegraphics[height=10cm]{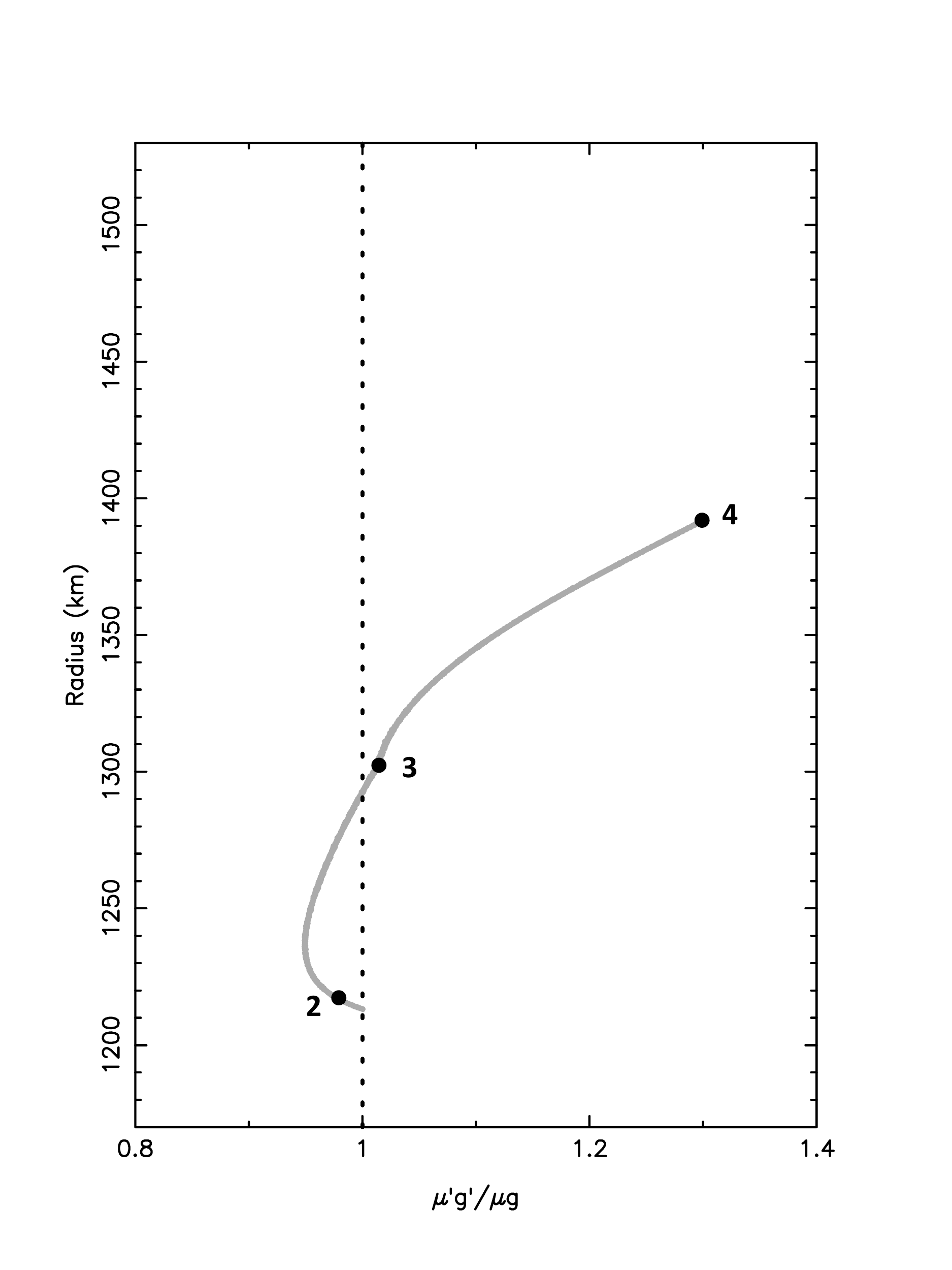}
\caption{%
\normalsize
Left panel - 
Solid line: our best temperature profile (see Fig.~\ref{fig_model}).
Gray line: an example of a prescribed profile with milder mesospheric thermal gradient,
here a 10~K drop between the stratopause and the mesopause.
Right panel -
The ratio $\mu' g'/\mu g$, as defined by Eq.~\ref{eq_mu_g}, corresponding to the 
gray, prescribed profile of the left panel. 
The points numbered 2, 3 and 4 correspond respectively to the 
stratopause, the inflexion point and the mesopause 
(see also Fig.~\ref{fig_model} and text for details).
}%
\label{fig_mu_g_r}
\end{figure*}

{\it Global Pluto's atmospheric model.}
We find that a unique thermal model can fit satisfactorily twelve light-curves observed
in 2012 and 2013  (Figs.~\ref{fig_fit_18jul12} and \ref{fig_fit_04may13}), assuming a spherically
symmetric and clear (no haze) atmoshere.
The parameters defining our best model are listed in Table~\ref{tab_atmo_fit} (see also Fig.~\ref{fig_model}), and
the various resulting profiles (density, temperature and thermal gradient) are displayed in 
Figs.~\ref{fig_T_r_invers_par12}, \ref{fig_T_r_invers_par12_bottom} and \ref{fig_T_r_invers}.
The absolute vertical scale of our global model has an internal accuracy of about $\pm 1$~km (Table~\ref{tab_atmo_fit}).
However, this error is amplified to $\pm 5$~km at the bottom of the profiles (Fig.~\ref{fig_T_r_invers_par12_bottom}), 
because of the uncertainty on the residual stellar flux (Fig.~\ref{fig_fit_18jul12_NACO}) in the central part of
the occultation observed by NACO on 18 July 2012.

We quantify in this work the propagation of the photometric noise into the 
density, temperature and thermal gradient profiles (Eqs.~\ref{eq_sigma_all} and Fig.~\ref{fig_T_r_invers_par12}).
The key parameter that governs the noise propagation is the radius $r_0$ in the amosphere at which 
the stellar drop caused by differential refraction is equal to the flux standard deviation. 
The radius $r_0$ can be estimated from Eq.~\ref{eq_nu0}, which 
includes all the quantities at work in a stellar occultation:
photometric noise, molecular refractivity, atmospheric scale-height and radius, and distance to the body.
For the NACO light-curve, we find $r_0= 1,565$~km, corresponding to a pressure level of about 14 nbar.

Although a satisfactory fit to all the data used here is provided by a unique model,
there are two slight, but significant departures from this global model. They are now discussed in turn.

{\it Pressure increase between 2012 and 2013.}
In the frame of our model (i.e. assuming a constant temperature profile), 
we detect a significant 6\% pressure increase (at the 6-$\sigma$ level) during the $\sim$9.5~months
separating the two events under study.
This means that Pluto's atmosphere was still expanding at that time, 
confirming the work of \citet{olk15}, which compiles and analyzes pressure measurements
between 1988 and 2013.

{\it Ingress/egress asymmetry of lower temperature profiles.}
Fig.~\ref{fig_fit_18jul12_NACO} shows that the stellar flux decreased from 
2.3\% to 1.8\% of its unocculted value during the central part
of the 18 July 2012 occultation, as observed by NACO from Paranal.
This corresponds to the primary stellar image first scanning the summer,
permanently lit Pluto northern hemisphere, and
then the winter low insolation southern hemisphere (Fig.~\ref{fig_XY}).
This confirms a similar trend pointed out by \cite{sic03} during another high SNR 
stellar occultation recorded in August 2002.
These authors interpreted this result as a surface boundary layer effect, 
where the lowermost scale-height adjusts itself to the surface temperature variegations,
which might explain the behavior displayed in Fig.~\ref{fig_T_r_invers_par12_bottom}.


Another interpretation of this trend is the  gradual entrance of the primary stellar image 
into an absorbing haze layer near Pluto's evening limb, a hypothesis that can be tested during
the  {\it New Horizons} flyby in July 2015.

{\it  Pluto's radius and density.} 
The extrapolation of our temperature profiles to the nitrogen saturation line implies that
nitrogen may condense at a Pluto's radius of $R_P=1,190\pm5$~km. 
However, the few kilometers above Pluto's surface remain {\it ``terra incognita"} as far as 
stellar occultations are concerned. In particular, the temperature gradients shown in 
Fig.~\ref{fig_T_r_invers_par12_bottom} may deviate from the simple extrapolation used here,
especially if haze layers affect the retrieved temperature profiles.
Although difficult to envisage because of the strong caustics that they cause, 
a troposphere below $1,190 \pm 5$~km cannot be excluded.
Combining high-resolution spectroscopic observations of gaseous methane,
combined with constraints from an occultation observed in 2002, \cite{lel09} conclude that
the troposphere depth cannot exceed about 17~km.
Consequently (and assuming that the temperature of the deep atmosphere did not change 
significantly since 2002), our observations constrain Pluto's radius to lie in the range 1,168-1,195~km.
More recently, combining constraints from spectra and a preliminary analysis 
of the occultation data presented in this work, 
\cite{lel15} concluded that Pluto should have a radius between 1,180-1,188 km,
some 2-8~km below the condensation radius of 1,190~km that we derive above.

From a Pluto's mass of $M_P = 1.304 \pm 0.006  \times$ 10$^{22}$~kg \citep{tho08},
we derive a density $\rho_P= (1.802 \pm 0.007)(R_P/1,200~{\rm km})^{-3}$ g cm$^{-3}$.
Our estimation $R_P=1,190\pm5$~km thus implies 
$\rho_P= 1.85 \pm 0.02$ g cm$^{-3}$ in the absence of troposphere, and 
a range $\rho_P= 1.83-1.95$ g cm$^{-3}$ if a troposphere is allowed.
This is larger, but not by much, than Charon's density, $\rho_C= 1.63 \pm 0.05$ g cm$^{-3}$ (Ibid.).

{\it The mesospheric negative thermal gradient.}
Pluto's stratopause occurs near 1,215~km (pressure $p= 6.0$ $\mu$bar), 
with a maximum temperature of $110 \pm 1$~K, 
where the error bar applies to the best inverted profile (NACO 18 July 2012), and
stems from the uncertainty on the Pluto + Charon flux contribution (Fig.~\ref{fig_T_r_invers_par12}).

Above the stratopause, and up to about 1,390~km, our best 2012 and 2013 occultation light-curves yield
inverted temperature profiles with a negative thermal gradient close to -0.2~K~km$^{-1}$ which amounts
to a total decrease of 30~K for the temperature between 1,215 and 1,390~km
(Figs.~\ref{fig_T_r_invers_par12},\ref{fig_T_r_invers})

Explaining this negative gradient by CO cooling requires a mixing ratio ($200 \times 10^{-4}$) 
that is too high by a factor of 40 compared to current measurements \citep{lel11}.
Cooling by HCN is also discussed in this paper. It appears to be a possible alternative solution, 
but only if it remains largely supersaturated in the mesosphere.

Changing the temperature boundary condition may suppress the negative gradient, 
but at the expense of creating a warm, unexplained thermal profile above 1,350~km.
We have investigated more exotic solutions, like zonal winds or compositional variations
that would  ``unbend" the retrieved temperature profiles, allowing a more isothemal mesosphere.
However, no realistic models could be built upon those alternative assumptions.
Again, the {\it New Horizons} flyby will provide constraints on the temperature boundary conditions
and atmospheric composition that will be used to discriminate between the various solutions
decribed here.

\vspace{0.5cm}

{\bf Acknowledgements.}
We acknowledge support from the French grants 
``Beyond Neptune"  ANR-08-BLAN-0177 and
``Beyond Neptune II"  ANR-11-IS56-0002.

A. Dias-Oliveira thanks the support of the following grants: CAPES (BEX 9110/12-7) FAPERJ/PAPDRJ (E-45/2013).

R. Vieira-Martins acknowledges the following grants: CNPq-306885/2013, CAPES/Cofecub-2506/2015, FAPERJ/PAPDRJ-45/2013.

TRAPPIST is a project funded by the Belgian Fund for Scientific
Research (F.R.S.-FNRS) under grant FRFC 2.5.594.09.F, with
the participation of the Swiss National Science Foundation.

J. L. Ortiz and N. Morales acknowledge funding from Proyecto de Excelencia de la Junta
de Andaluc\'ia, J.A. 2012-FQM1776 and from FEDER funds.

M. Gillon and E. Jehin are F.R.S.-FNRS Research Associates.

C. Opitom acknowledge the support of the F.R.S.-FNRS for her PhD thesis.

R. Leiva is supported by CONICYT PCHA/Doctorado Nacional scholarship.

M. Assafin thanks the CNPq (Grants 473002/2013-2 and 308721/2011-0) and FAPERJ (Grant E-26/111.488/2013). 

A. R. Gomes-J\'unior thanks CAPES. 

We also thank Caisey Harlingten, for the use of his 20 inch telescope in San Pedro de Atacama.

\vspace{0.5cm}

\bibliography{mnemonic,biblio}   

\newpage

\newpage


\clearpage

\clearpage

\begin{table*}[!htb]
\tabletypesize{\scriptsize}
\caption{Circumstances of the 18 July 2012 Pluto occultation}
\label{tab_18jul}
\begin{tabular}{*{5}{l}}
\hline\hline
Site                       & Lat. (d:m:s)  & Telescope               &  Exp. time/cycle (s)\footnotemark[1]          &      Observers       \\
                              & Lon. (d:m:s) & Instrument/filter       &                                           &                      \\
                             & alt  (m)          &                                      &                                        &                              \\
\hline\hline
Observatory UC     & 33:16:09.0 S   & 0.4 m               &  1.0/1.0                  & R. Leiva Espinoza   \\
(Santa Martina)    & 70:32:04.0 W  & CCD/clear     &                                 &                                     \\
                              & 1,450                &                           &                               &                        \\
\hline
Cerro Burek          & 31:47:12.4 S  & ASH\footnotemark[2] 0.45 m      &  13.0/15.7            & N. Morales           \\
                               & 69:18:24.5 W  & SBIG-STL11000/clear      &                               &                       \\
                              & 2,591                 &                           &                               &                        \\
\hline
Paranal                & 24:37:31.0 S   & VLT Yepun 8.2 m  &  0.2/0.2           &   J. Girard         \\
                              & 70:24:08.0 W  & NACO/H                 &                         &                            \\
                             & 2,635                &                                  &                         &                            \\
\hline
San Pedro        & 22:57:12.3 S   & ASH2  0.4 m    &  13.0/15.44      & N. Morales           \\
de Atacama     & 68:10:47.6 W  & SBIG-STL11000/clear        &                           &                       \\
                         &   2,397             &                            &                           &             \\
\hline
Huancayo                 & 12:02:32.2 S   &   0.2 m               &  10.24/10.24\footnotemark[3]            & E. Meza           \\
                         & 75:19:14.7 W  & CCD/clear        &   5.12/5.12\footnotemark[3]              &                       \\
                         & 3,344                &                            &                           &             \\
\hline\hline
\end{tabular}
~\\
\it Notes. \rm 
\footnotemark[1]{Cycle is defined as the exposure time plus the readout time also known as dead time. Observations with the same exposure time and cycle have no dead time.}

\footnotemark[2]{ASH - Astrograph for the Southern Hemisphere.}

\footnotemark[3]{Exposure time was changed at  04:11:46 UT}
\end{table*} 

\clearpage

\begin{table*}
\tabletypesize{\scriptsize}
\caption{Circumstances of the 04 May 2013 Pluto occultation}
\label{tab_04may}
\begin{tabular}{lllll}
\hline\hline
Site                   & Lat. (d:m:s)    & Telescope                   &  Exp. time/cycle (s)     &      Observers       \\
                          & Lon. (d:m:s)   & Instrument/filter           &                       &                               \\
                         & alt  meters       &                                      &                                        &                              \\
\hline\hline
Cerro Burek    &  31:47:12.4 S   & ASH 0.45m               &  6/8  & J.L. Ortiz \\
                         &  69:18:24.5 W  &  SBIG-STL11000/clear              &         &  N. Morales \\
                         & 2,591 m             &                                  &          &  \\
\hline

CASLEO         &  31:47:55.6 S    &   Jorge Sahade 2.15m  &  5/6.8    & R. Gil-Hutton \\
(Leoncito)        & 69:17:44.9 W    &  CCD/R                          &                          & C. Lopez-Sisterna \\
                         & 2,492   m            &                                         &                          &                                   \\
\hline
Cerro Tololo   & 30:10:03.4  S      & PROMPT\footnotemark[1] 0.4m   & 5/8                               &   J. Pollock \\
                         & 70:48:19.0  W    & P1, P3, P4, P5    &  P3 offset 2 s          &                      \\
                         & 2,207             & CCD/clear          &  P4 offset 4 s          &                      \\
                         &                                &                               &  P5 offset 6 s          &                      \\
\hline
%
%
La Silla           &  29 15 21.276 S   & Danish 1.54m                                             & 0.1/0.1          & L. Mancini         \\
                        & 70 44 20.184  W  &  Lucky Imager/Z  ($\lambda >$ 650nm   & several          &                            \\
                        &  2,336           & CCD/iXon response)         & interruptions               &                            \\
                        &                  &                            &  due to image  &   \\
                        &                  &                            &  cube writing  &   \\
\hline
La Silla           & 29 15 16.59 S    & TRAPPIST\footnotemark[2] 0.6m   & 4.5/6                   & E. Jehin,                                \\
                        & 70  44 21.82  W  & CCD/clear             &                             &  A. Decock, M. Gillon         \\
                        & 2,315                     &                                 &                            & C. Opitom                            \\
\hline
Pico dos Dias & 22 32 07.8 S       &  B\&C\footnotemark[3] 0.6m          & 5/5.40              &  M. Assafin,                      \\
                          & 45  34 57.7  W    &  CCD/I                 &                           & A. Ramos-Gomes Jr  \\
                         & 1,811                      &                             &                           &       \\
\hline
Ponta Grossa & 25 05 24.00 S     &  Meade 16 0.4m            & 5               &  M. Emilio        \\
                          & 50 09 36.00 W    & CCD/clear                    &     Tecnichal    &            \\
                          & 909                       &                               & Problems  &             \\
\hline 
Cerro Paranal & 24:37:31.0   S      & UT4 Yepun 8.2m  & 0.2/0.2                   &  G. Hau     \\
                           & 70:24:08.0  W     & NACO/H                 &                               &                    \\
                           & 2,635                   &                                  &                               &                    \\
\hline
San Pedro      &   22:57:12.3 S       & Caisey 0.5m f/8          & 3/4.58         & A. Maury \\
de Atacama   &   68:10:47.6 W       & CCD/V  &               &                   \\
                         &   2,397                    &                                  &                     &                   \\
 \hline
                        &    --                           & Caisey 0.5m f/6.8   & 4/4.905   & L. Nagy \\
                        &                                 & CCD/B                     &                  &                \\
\hline
                       &   --                            & CAO\footnotemark[4]   0.4m                  & 4/6.35     & J.F. Soulier \\
                       &                                & CCD/R                        &                 &                      \\
\hline
                        &  --                            & ASH2   0.4m              & STL11000                 & N. Morales \\
                        &                                 &                                      & technical problem   &                      \\
 \hline
                       &   --                            & OPSPA\footnotemark[5] 0.3m                 &  5/11.1          & J.  Fabrega Polleri \\
                        &                                 & CCD/clear                     &                      &                \\
\hline
Huancayo                &   12:02:32.2 S      & Meade 8 0.2 m                 &  10.24/10.24     & E. Meza  \\
                        &   75:19:14.7 W      & CCD/clear             &  Negative chord   &                \\
                         &   3,344            &                       &                   &                   \\
\hline\hline
\end{tabular}
~\\
\it Notes. \rm 
\footnotemark[1]{PROMPT: Panchromatic Robotic Optical Monitoring and Polarimetry Telescopes.}

\footnotemark[2]{\cite{jeh11}}.
\footnotemark[3]{B\&C: Boller \& Chivens.}
\footnotemark[4]{CAO: Campo Catino Observatory.} \\
\footnotemark[5]{OPSPA:  Observatorio Paname\~no en San Pedro de Atacama.} \\
\end{table*}

\clearpage

\clearpage

\clearpage

\clearpage

\clearpage

\clearpage

\clearpage

\clearpage

\clearpage

\clearpage

\begin{table*}
\centering
\caption{Parameters of the two occultations}
\label{tab_param}
\begin{tabular}{lll}
\hline\hline
& 18 July 2012 &  04 May 2013 \\
\hline\hline
Star coordinates\footnotemark[1]                    &  $\alpha$= 18h 32m 14.6720s &  $\alpha$= 18h 47m 52.5322s \\
                                                                             &  $\delta$= -19d 24' 19.295"      &  $\delta$= -19d 41' 24.3738" \\
Ephemeris                                                          & DE413/PLU022  & DE413/PLU031  \\
Pluto geocentric distance                                &  $4.68244 \times 10^9$ km (at 04:13 UT) &  $4.76882 \times 10^9$ km (at 08:23 UT) \\
Sub-observer and sub-solar latitudes\footnotemark[2]   & B= +47.10d, B'= +47.54d & B= +49.95d, B'= +48.64d \\
Pluto's north pole position angle\footnotemark[2]            & P=  -56.88d & P=  -52.91d \\
Shadow velocity                                               &  $\approx$ 23.0 km s$^{-1}$ &  $\approx$ 10.6 km s$^{-1}$ \\
Magnitudes\footnotemark[3]                           &  V= 14.7, R=13.7, K= 10.9 & V= 14.1, R=14.0, K= 12.4 \\
\hline\hline
\end{tabular}
~\\
~\\
\raggedright
\footnotemark[1]J2000, UCAC2 system.

\footnotemark[2]Assuming the Pluto's north pole position (J2000) of \cite{tho08}: 
$\alpha_{\rm p}$= 08h  52m 12.94s, 
$\delta_{\rm p}$= -06d 10' 04.8" \\
\footnotemark[3]{From NOMAD catalog \citep{zac04}}
\end{table*}

\begin{table*}
\centering
\caption{Pluto atmospheric model}
\label{tab_atmo_fit}
\begin{tabular}{lll}
\hline\hline
\multicolumn{3}{c}{Physical parameters} \\
\hline
Pluto's mass\footnotemark[1]                               & \multicolumn{2}{l}{$GM= 8.703 \times 10^{11}$ m$^3$ s$^{-2}$} \\
Nitrogen molecular mass\footnotemark[2]         & \multicolumn{2}{l}{$\mu= 4.652 \times 10^{-26}$ kg} \\
Nitrogen molecular refractivity\footnotemark[3] & \multicolumn{2}{l}{$K = 1.091 \times 10^{-23} 
                                                                                  + (6.282 \times 10^{-26}/\lambda_{\rm \mu m}^2)$ cm$^3$ molecule$^{-1}$}   \\
Boltzmann constant                                              &  \multicolumn{2}{l}{$k= 1.380626 \times 10^{-23}$ J K$^{-1}$} \\
\hline\hline
\multicolumn{3}{c}{The nine free parameters of the best temperature profile\footnotemark[4]} \\
\hline
$r_1$, $T_1$, $dT/dr(r_1)$   & \multicolumn{2}{l}{$1,190.4 \pm 1$ km, 36 K, 16.9 K km$^{-1}$}  \\
$r_2$, $T_2$                           & \multicolumn{2}{l}{$1,217.3$ km, 109.7 K} \\
$r_3$, $T_3$                           & \multicolumn{2}{l}{$1,302.4$ km, 95.5 K (implying $dT/dr(r_3)= -0.206$ K km$^{-1}$)}  \\
$r_4$, $T_4$                           & \multicolumn{2}{l}{$1,392.0$ km, 80.6 K} \\
\hline
$c1,c2$  & \multicolumn{2}{l}{$1.41397736 \times 10^{-3}$, $2.59861886 \times 10^{-3}$} \\
$c3,c4$  & \multicolumn{2}{l}{$-2.19756021 \times 10^{-6}$, $-4.81764971 \times 10^{-7}$} \\
$c5, c6$ & \multicolumn{2}{l}{$8.66619700 \times 10^{-8}$, $-3.6213609  \times 10^{4}$} \\
$c7,c8$  & \multicolumn{2}{l}{$8.2775269   \times 10^{1}$, $-6.27372563 \times 10^{-2}$} \\ 
$c9$       & \multicolumn{2}{l}{$1.58068760 \times 10^{-5}$} \\
\hline\hline
\multicolumn{3}{c}{The three free parameters particular to each event\footnotemark[5]} \\
\hline
 & 18 July 2012 & 04 May 2013 \\
\hline
Pressure at $r= 1,275$~km, $p_{1,275}$   & $2.16 \pm 0.02$ $\mu$bar   & $2.30 \pm 0.01$ $\mu$bar \\
Time of closest geocentric approach          &  04:13:37.24$\pm 0.07$ UT & 08:22:27.11$\pm 0.09$ UT  \\   
Distance of closest geocentric approach\footnotemark[6]  &  $-404.6 \pm 2.7$~km &  $-723.5\pm 2.7$~km \\
\hline\hline
\end{tabular}
~\\
~\\
\raggedright
\footnotemark[1]{\cite{tho08}.}
\footnotemark[2]{Assumed to be the only constituent in the ray tracing code, see text.}
\footnotemark[3]{\cite{was30}. For both NACO observations of 2012 and 2103, H band ($\lambda=1.6$~$\mu$m)} was used.
\footnotemark[4]{Or equivalently, the nine coefficients $c1,...c9$,
see text and Fig.~\ref{fig_model} for the definition of the various quantities given here.}
\footnotemark[5]{So that there are twelve free parameters for each date.}
\footnotemark[6]{Distance of Pluto's center to the star at closest approach, projected in the sky plane,
as seen from the geocenter. A negative value means that Pluto's center went south of the star in the sky plane.}
\end{table*} 



\appendix

\section{Synthetic light curves}

\subsection{Parametrized temperature profile}
We define a parametric model for Pluto's atmosphere temperature profile, $T(r)$,
where $r$ is the radius, i.e. the distance to Pluto's center.
The model must be 
detailed enough to capture the main features revealed by the inversions (Figs.~\ref{fig_T_r_invers} and \ref{fig_model}),
but still simple enough to allow 
an easy and meaningful control of $T(r)$ and
an assessment of the error bars associated with each parameter.
The features we want to describe are: 
$(i)$ a thin stratosphere just above the surface, with a strong increase of temperature with altitude, 
$(ii)$  an ``elbow" where the temperature reaches its maximum, marking the stratopause, 
$(iii)$ an intermediate region with a mild negative gradient, and finally 
$(iv)$ an isothermal upper branch.

These features define three regions, bounded by four points 1,...4 at prescribed radii $r_1,...r_4$, 
see Fig.~\ref{fig_model}. 
More precisely, the profile $T(r)$ is generated as follows:
\begin{equation}
 \left\{
  \begin{array}{lll}
  \displaystyle
  c1 \cdot r + c2 \cdot T(r) + c3  \cdot r  T(r) + c4 \cdot  r^2 + c5 \cdot T^2(r) =1 & \textrm{for} & r_1 \leq r \leq r_3 ~~\textrm{(hyperbolic branch)} \\
  & & \\  
  T(r) = c6 + c7 \cdot r + c8 \cdot r^2 + c9 \cdot r^3                                                & \textrm{for} & r_3 \leq r \leq r_4 ~~\textrm{(polynomial branch)} \\
  & & \\
  T(r)= T_{\rm iso}                                                                                                       & \textrm{for} & r_4 \leq r     ~~\textrm{(straight line)}           \\
  \end{array}
 \right.
 \label{eq_T_r_model}
\end{equation}

Note that $r_2$ does not appear in the equations above, and 
is defined as the radius where the temperature reaches its maximum (Fig.~\ref{fig_model}).
The functional forms chosen here (hyperbolic, polynomial and straight lines)
are {\it not}  based on physical grounds, but rather, are
empirical and simple formulae that satisfactorily fits the observed profiles 
(Fig.~\ref{fig_T_r_invers_par12}). 

The  parameters $c1,...c9$ are determined to ensure that $T(r)$ is continuous both
in temperature and its derivative, $dT/dr$, at points 1, 3 and 4.
Those conditions provide algebraic systems that are solved by 
a classical Gauss-Jordan method \citep{pre92}.

In practice:
(1) 
we fix the temperature $T_1= T(r_1)$ at the bottom of the profile, 
together with its gradient $(dT/dr)_1$.
%
(2) 
We fix the value of 
the maximum of temperature $T_2= T(r_2)$ at $r_2$ and 
the temperature $T_3= T(r_3)$ at the inflexion point 3.
We thus have three boundary conditions for $T$: $T_1$, $T_2$, $T_3$ at $r_1$, $r_2$ and $r_3$, respectively, 
and two boundary conditions for $dT/dr$: $(dT/dr)(r_1)= (dT/dr)_1$ and $(dT/dr)(r_2)=0$, 
which fixes the five coefficients $c1,...c5$. 
Note in passing that the values of $c1,...c5$ then impose the temperature gradient $(dT/dr)_3$ at $r_3$; 
(3) 
We fix the temperature $T_{\rm iso}$ at $r_4$, the point where the isothermal branch is reached. 
This provides two boundary conditions in $T$: $T_3$ and $T_{\rm iso}$ at $r_3$ and $r_4$, respectively,
plus two boundary conditions for $dT/dr$: $(dT/dr)_3$ at $r_3$ and $(dT/dr)(r_4)= 0$, 
thus fixing the remaining four coefficients $c6...c9$. 

The locations of points 1, 2, 3 and 4 in the space $(T,r)$  are chosen to best fit the observed profiles, see the main text for details.
Once $T(r)$ is defined, the gas number density profile $n(r)$ is obtained by integrating
the first order differential equation:
\begin{equation}
\frac{1}{n} \cdot \frac{dn}{dr} =  -\left[   \frac{\mu g(r)}{kT} +  \frac{1}{T} \cdot \frac{dT}{dr}  \right],
\label{eq_n_r}
\end{equation}
derived from the equation of state for 	an ideal gas, and the hydrostatic equation.
Here,
\begin{equation}
g(r)=\frac{GM}{r^2}
\label{eq_g_r}
\end{equation}
 is the acceleration of gravity, assuming a spherical, homogeneous planet.
The values of $\mu$ (Table~\ref{tab_atmo_fit}) corresponds to molecular nitrogen, 
assumed to be the unique gas present in the atmosphere.
Also listed in Table~\ref{tab_atmo_fit} are 
the Boltzmann constant $k$ and Pluto's mass parameter $GM$.

A  boundary condition is required to integrate Eq.~\ref{eq_n_r}, 
e.g.  the pressure $p_{\rm 1,275}$ at $r= 1,275$~km, which
fixes the needed boundary condition $n_{\rm 1,275} = P_{\rm 1,275}/kT_{\rm 1,275}$.
Finally, the refractivity $\nu(r)$ of the gas (index of refraction minus unity) is given by 
\begin{equation}
\nu(r)= K \cdot n(r), 
\label{eq_nu_r}
\end{equation}
where the molecular refracticity is given in Table~\ref{tab_param}, 
assuming again pure molecular nitrogen. 
Once $\nu(r)$ is obtained, we can derive the vertical refractivity gradient $d\nu/dr$ 
that is used in the ray tracing code, see below.

The inversions proceed the other way around: 
the light-curves provide $d\nu/dr(r)$ through an abelian integral \citep{vap73}, 
then $\nu(r)$, from which $n(r)$ is derived (Eq.~\ref{eq_nu_r}),  
followed by the temperature profile, once a boundary condition is given for $T$ (Eq.~\ref{eq_n_r})

\subsection{Ray tracing}

For small values of $\nu$ (as it is the case here) and under spherical symmetry assumption,  
a stellar ray is deviated by
$d\omega= (\partial \nu/\partial r) \cdot ds$ 
(Snell-Descartes law) as it moves along an elementay path $ds$. 
%
%
In principle, a ray tracing code should account for the curvature of the stellar ray as it is
refracted in the atmosphere.
In practice, however, it is enough to assume that the ray has a rectilinear trajectory in the entire atmosphere.
In fact, the maximum total deviation $\omega$ suffered by the ray is very small for ground-based occultations, 
more precisely of the order of Pluto's apparent angular radius, $\sim$0.05~arcsec, 
so that $\omega < \sim3 \times 10^{-7}$ rad.
Most of that deviation occurs in the deepest scale height $H$ traversed at radius $r$, 
which represents a traveled length of $l \sim \sqrt{2\pi r H}$ \citep{bau53}.
Taking typical values of  $r \sim 1,200$ km and $H < \sim 50$ km, we get $l < \sim 600$ km,
i.e. a deviation inside the atmosphere of $\sim \omega \cdot l < 0.2$~meters, which is negligible compared to the 
scales probed by ground-based stellar occultations. 

The numerical integration of Eq.~\ref{eq_n_r}, using a second order scheme, 
provides $n(r_i)$ at discreet layers of radii $r_i$, from which the 
refractivity $\nu_i$ and its gradient $(d\nu/dr)(r_i)$ are calculated. 
The total deviation along the straight line $s$ is then:
\begin{equation}
\omega = \sum_{i} \Delta \omega_i= \sum_{i} (d\nu/dr)(r_i)\cdot \Delta s_i,
\label{eq_sum}
\end{equation}
where $\Delta s_i$ is the path along $s$ traveled inside the layer $i$.
Then, for a closest approach $r$ of a ray to Pluto's center, the corresponding distance $z$
to the shadow center upon arrival on Earth is 
\begin{equation}
z= r + \omega(r) \cdot D,
\label{eq_z}
\end{equation}
where $D$ is Pluto's geocentric distance.
The observed stellar flux is then
\begin{equation}
\Phi(z)= f \frac{dr}{dz}= \frac{1}{1+D d\omega/dr},
\label{eq_focal}
\end{equation}
where $f= r/z$ is the focusing factor due to the (assumed circular) limb curvature, see \cite{sic99}.

The thickness $\Delta r_i$ of the individual refracting layers has been adjusted to 30~meters to
minimize numerical noise, while keeping computing times reasonably low.
Similarly, the sampling for $r$ (the closest distance of the rays to Pluto's center) has been adjusted 
so that adjacent rays arrive at separation $\Delta z \approx$ 1 km in the shadow plane.

Once the table $(r,z,\Phi(z))$ has been completed, the synthetic flux at a given site and given moment
(corresponding to a distance $z_{\rm obs}$ of the observer to the shadow center) is calculated by interpolation. 
If several stellar images are present, all the fluxes are summed. 
In the particular case of a spherically symmetric atmosphere, and 
for a given distance $z_{\rm obs}$, there is 
a primary image corresponding to $z= z_{\rm obs}$, and a 
secondary image corresponding to $z= -z_{\rm obs}$.

\begin{figure}[!h]
\vspace{00pt}
\centerline{
\includegraphics[height=70mm,trim=0 0 40 0]{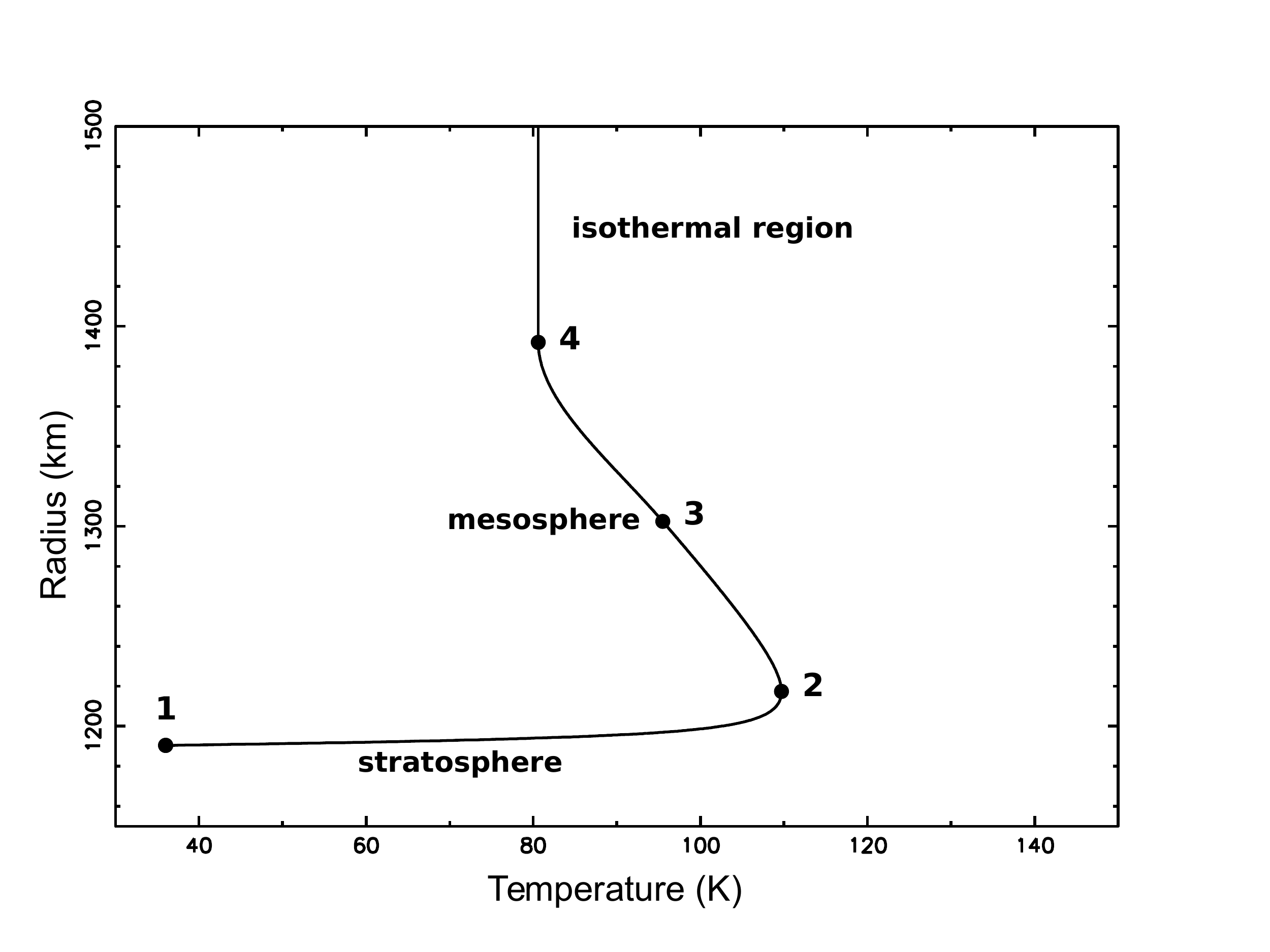}
\includegraphics[height=70mm,trim=40 0 0 0]{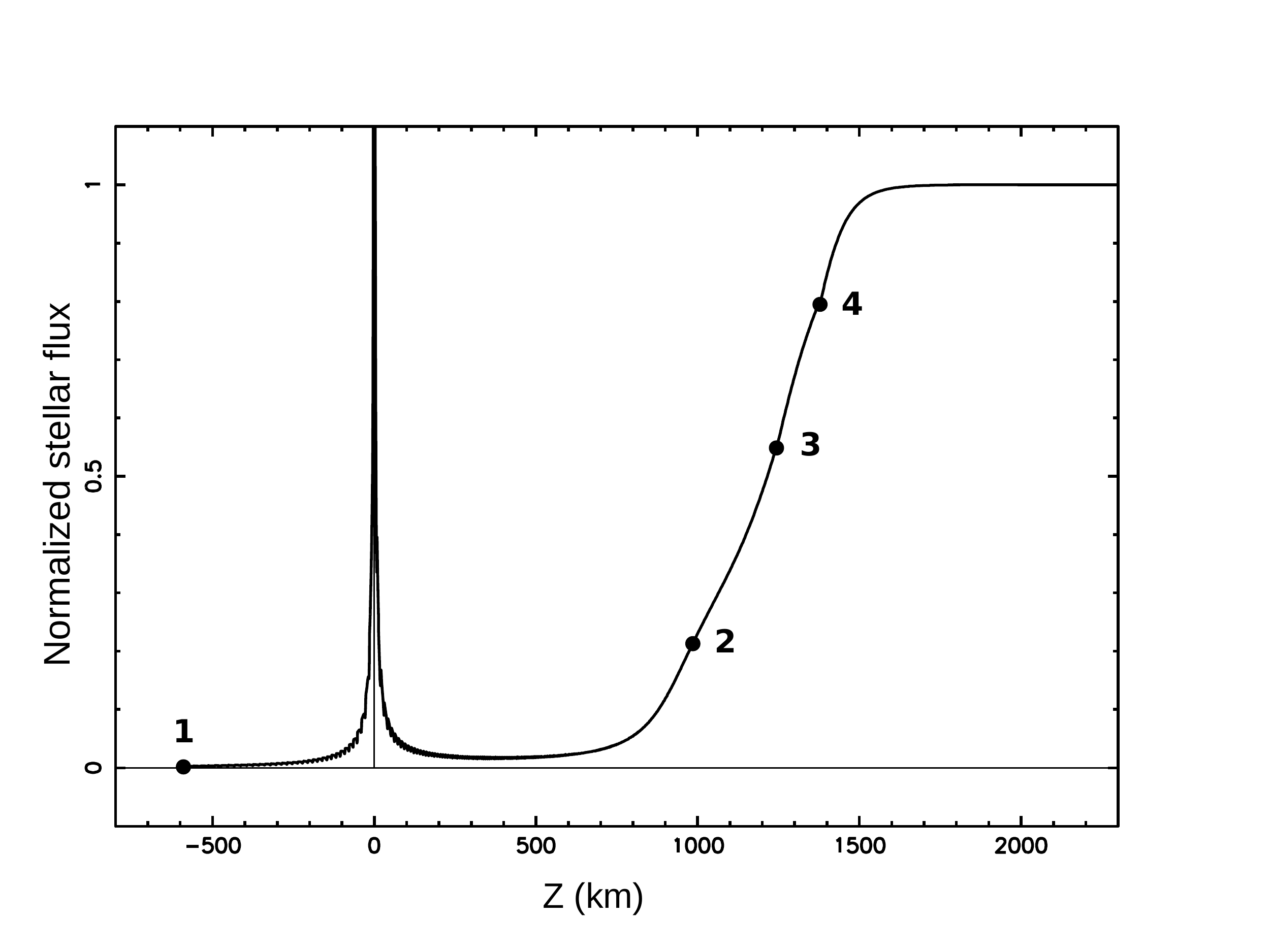}
}%
\caption{%
\normalsize
Left -
The temperature profile $T(r)$ that best fits our 2012 and 2013 light-curves, 
see Figs.~\ref{fig_fit_18jul12} and \ref{fig_fit_04may13}.
The parameters used to generate this profile are given in Table~\ref{tab_atmo_fit}.
Total thickness of the atmosphere: 1100 km,
vertical sampling of the model:  0.03 km.
Right -
The corresponding synthetic flux in the shadow plane for the 18 July 2012 occultation. 
Here, $z$ is the distance to the shadow center, 
with the four points corresponding to those of the left panel.
}%
\label{fig_model}
\end{figure}

The lowest radius $r_1$ considered in the model (1,190.4~km, see Table~\ref{tab_atmo_fit}  and Fig.~\ref{fig_model})
is adjusted so that the corresponding flux received in the shadow plane is $\sim 10^{-3}$ of the
unocculted stellar flux, negligible compared to the noise level of the best light-curves. 
The upper limit for the atmosphere has been fixed to a radius of about 2,300~km. 
This corresponds to a pressure level of about 0.05~nbar, 
at which point the stellar drop is several orders of magnitudes less than the noise in our best light-curves.

The profile that best fits our light-curves is shown in Fig.~\ref{fig_model}.
The trajectories of the primary and secondary stellar images as seen from VLT on 18 July 2012
are displayed in Fig.~\ref{fig_XY}.

The best fits to the observed light-curves are shown in Figs.~\ref{fig_fit_18jul12} and \ref{fig_fit_04may13}. 
Their quality is assessed through the so-called $\chi^2$ per degree of freedom:
\begin{equation}
\chi^2_{\rm dof}= \frac{\chi^2}{N-M } = \frac{1}{N-M }\sum_{i=1}^{N} \left(\frac{\Phi_{\rm obs,i}-\Phi_{\rm syn,i}}{\sigma_i}\right)^2,
\label{eq_chi2}
\end{equation}
where 
$\Phi_{\rm obs,i}$ (resp. $\Phi_{\rm syn,i}$) is the observed (resp. synthetic) stellar flux of the $i^{\rm th}$ data point,  
$\sigma_i$ is its associated standard deviation, 
where the summation is extended to the $N$ data points from all the light-curves used in the fit, and
$M$ is the number of free parameters of the model.
As we have nine coefficients $c_1,...c_9$ to define $T(r)$ (Eq.~\ref{eq_T_r_model}), 
a boundary condition $p_{\rm 1,275}$ and 
two quantities to define Pluto's center, $M=12$.

\section{Noise propagation}

\subsection{Photometric noise}

Here we estimate the effect of photometric noise in an occultation  light-curve 
on the retrieved density, temperature and temperature gradient profiles. 
We denote $\delta$ the fluctuation of a given quantity, and $\sigma= 	\sqrt{\bar{\delta^2}}$ its 
standard deviation, where the bar denotes average values.
For estimation purposes, it is enough to assume here (but not in the ray-tracing or inversion procedures)
that  the atmosphere has locally a constant 
density scale height $H$ that is small compared to the planet radius.  
The stellar flux is then given by the Baum and Code (BC) equation \citep{bau53}: 
\begin{equation}
\frac{1}{\Phi} - 2 + \log\left(\frac{1}{\Phi}-1\right)= -\frac{z-z_{1/2}}{H},
\label{eq_BC}
\end{equation}
where $z_{1/2}$ is the position in the shadow plane at which $\Phi= 1/2$ (half-light  level).

We focus on the top of the profiles, corresponding to $\Phi \sim 1$, 
so that Eq.~\ref{eq_BC} becomes $\Phi(z) \sim 1 - \exp[-(z-z_{1/2})/H]$.
Morever, for $\Phi \sim 1$, the stellar ray deviation $\omega$ is small, and 
we can equate $r$ and $z$, see Eqs.~\ref{eq_z} and \ref{eq_focal}, where $f \sim 1$.
In the BC approximation, we have  $\omega \sim - \nu \sqrt{2\pi r/H}$, where $\nu$ is the refractivity at $r$.
As the atmosphere density profile is basically exponential, $d\omega/dr \sim  -\omega/H =  \nu \sqrt{2\pi r/H^3}$,
so that Eq.~\ref{eq_focal} can be used to estimate the expected refractivity corresponding to a stellar flux $\Phi$, namely
$\nu \sim (1-\Phi) \sqrt{H^3/2\pi r D^2}$

We denote $\nu_0$ and $r_0$ the refractivity and corresponding radius where
the stellar drop is equal to the standard deviation of the flux, $\sigma_\Phi$, i.e.
\begin{equation}
\nu_0 \sim \sigma_\Phi  \sqrt{\frac{H^3}{2\pi r D^2}}.
\label{eq_nu0}
\end{equation}

Thus, $r_0$ is the radius where the stellar drop starts to be barely significant, 
given the photometric noise.
At the upper part of the profiles, we have $H \sim 60$~km. 
The 18 July 2012 NACO lightcurve has a photometric standard deviation of $\sigma_\Phi= 0.011$.
Using the value of $D$ given in Table~\ref{tab_param}, we obtain $\nu_0 \sim 1.3 \times 10^{-11}$.
Assuming a pure $N_2$ atmosphere, we obtained the corresponding molecular density
$n_0 = \nu_0/K \sim 6 \times 10^{12}$~cm$^{-3}$, 
which is reached at radius 
$$
r_0 \sim 1,565~{\rm km}.
$$
For $\Phi \sim 1$ (and $f \sim 1$), and using the results above, Eq.~\ref{eq_focal} provides
$\Phi \sim 1 - D d\omega/dr = 1 + \sqrt{2\pi r D^2/H} (d\nu/dr)$.
For a \it noise-free \rm light-curve, we expect $\Phi = 1 - \sigma_\Phi \exp[-(r-r_0)/H]$.
In reality, $\Phi$ is affected by fluctuations $\delta_\Phi$, so that the retrieved refractivity gradient is in fact:
$d\nu/dr= (\nu_0/H) \cdot [-\exp[-(r-r_0)/H] + \delta_\Phi/\sigma]$.
Consequenly, the standard deviation associated with each point of the $(d\nu/dr)(r)$ profile 
(and restricting ourselves to the top of the profile) is: 
$$
\sigma_{d\nu/dr} \sim \frac{\nu_0}{H}
$$

The profile $(d\nu/dr)(r)$ is the primary result derived from the light-curve, 
and from which all the other profiles are deduced.
Once $d\nu/dr$ is known, we have to estimate
$\nu(r)= \nu_1 + \int_{r_1}^{r} (d\nu/dr)dr$, where $(r_1,\nu_1= \nu(r_1))$ is a boundary condition. 
The integration is performed numerically by taking 
$\nu(r)=  \nu(r_1) + \sum_{i=1}^{N} (d\nu/dr)(r_i) \cdot \Delta r$, where 
$\Delta r$ is the spatial sampling of the data 
(i.e. $\Delta r=$ the star velocity perpendicular to the limb multiplied by the exposure time).
Thus, $r_i= r_1 + (i-1)\Delta r$ and $N= |r-r_1|/\Delta r$.
Adding the variances associated with individual $(d\nu/dr)(r_i)$'s, we obtain:
$$
\sigma_\nu \sim \sqrt{\frac{|r-r_1|}{H} \frac{\Delta r}{H}} \nu_0 \sim \sqrt{\frac{\Delta r}{H}} \nu_0,
$$
where the second approximation stems from the fact that $r_1$ is chosen close to $r_0$
and that we are considering here the few top scale heights of the profiles, so that $|r-r_1|/H \sim {\cal O}(1)$.
Note that $\sigma_\nu=0$ for $r = r_1$. This is because $(r_1,\nu_1)$ is an arbitrary boundary condition,
and such, has no associated error bars.

From $n= \nu/K$, we obtain the standard deviation associated with 
the density gradient and the density itself:
$\sigma_{dn/dr}= \sigma_{d\nu/dr}/K$ and
$\sigma_n= \sigma_\nu/K$.
Moreover, from Eq.~\ref{eq_n_r}, and assuming an isothermal upper atmosphere, we obtain 
$\delta_{dT/dr}= -(T/n) \delta_{dn/dr}$, so that $\sigma_{dT/dr}= (T/n) \sigma_{dn/dr}$. 
Finally, the temperature profile is obtained from the
numerical integration of $T(r)= T_2 + \int_{r_2}^r (dT/dr)dr$, where $(r_2,T_2= T(r_2))$ is an arbitrary boundary condition.
Using the same line of reasoning  as for $n(r)$, we obtain $\sigma_T$ by adding the variances 
$\sigma^2_{dT/dr}$ of all the points $i=1...N$ involved in the integration, where now $N=|r-r_2|/\Delta r$. 
Combining the results above, we obtain the following standard deviations for $n(r)$, $T(r)$ and $(dT/dr)(r)$:
\begin{equation}
\begin{array}{lll}
\displaystyle
\sigma_n \sim \sqrt{\frac{\Delta r}{H}} n_0,  &  
\displaystyle
\sigma_T 	\sim  T \sqrt{\frac{\Delta r}{2H}}  
\sqrt{\displaystyle \left|e^{2(r-r_2)/H}-1\right| } 
e^{(r_2-r_0)/H}, & 
\displaystyle
\sigma_{dT/dr} \sim \frac{T}{H}  e^{(r-r_0)/H}    \\
\end{array}
\label{eq_sigma_all}
\end{equation}

Fig.~\ref{fig_T_r_invers_par12} shows the $\pm 1\sigma$ envelopes at the upper parts of the various profiles. 
We take here $r_2= 1,390$~km, the radius at which we fix a prescribed temperature $T_2 \sim 81$~K.
Note again that $\sigma_T=0$ at $r=r_2$, as $(r_2,T_2)$ is an arbitrary boundary condition.
Finally, the envelopes $\pm 1\sigma$ are plotted only down to the half-light level ($r \sim 1,290$~km),
as the estimations made here apply only for the upper part of the light-curve.
In any case, below that level, the uncertainties in the profiles are dominated by the uncertainty
on the background Pluto + Charon contribution, see below.

\subsection{Effect of the Pluto and Charon flux contributions}

The stellar flux reaches its minimum value in the shadow at typically $z_{\rm min} \sim (z_{1/2})/2$, i.e. 
half-way between the half-light level and the shadow center, where the central flash occurs (Fig.~\ref{fig_model}). 
At the minimum, we have from Eq.~\ref{eq_BC}:  $\Phi_{\rm min} \sim  H/(z_{1/2} - z_{\rm min}) \sim 2H/z_{1/2}$.
Eq.~\ref{eq_n_r} then provides 
\begin{equation}
H = \left|\frac{n}{dn/dr}\right| = \frac{T}{(\mu g/k) +   (dT/dr)} \sim \frac{z_{1/2}}{2} \Phi_{\rm min}.
\label{eq_H_bottom}
\end{equation}

At the bottom of the temperature profile (stratosphere), $\mu g/k$ and $dT/dr$ are of same order of magnitude.
Consequently, increasing the value of the Pluto + Charon contribution to the light-curve decreases the
value of $\Phi$ (Fig.~\ref{fig_fit_18jul12_NACO}), thus increasing the retrieved gradient $dT/dr$. 
This is illustrated in Fig.~\ref{fig_T_r_invers_par12_bottom}.

\subsection{Effect of initial conditions}

Once the density profile $n(r)$ is derived from the inversion, 
Eq.~\ref{eq_n_r} yields the temperature profile $T(r)$,
provided a boundary condition $T_b=T(r_b)$ is fixed at an arbitrary level $r_b$.
Let us consider two possible solutions $T(r)$ and $(T+\Delta T)(r)$ that differ by $\Delta T(r_b)$ at $r_b$,
then expanding Eq.~\ref{eq_n_r} to first order in $(\Delta T/T)(r)$, we obtain:
$$
\frac{d}{dr} \left(\frac{\Delta T}{T}\right) \sim \frac{1}{H}  \left(\frac{\Delta T}{T}\right),
$$
where we have approximated $H \sim kT/\mu g$. 
Thus, as $r$ increases, the relative difference $\Delta T/T$ diverge exponentially as:
\begin{equation}
\left( \frac{\Delta T}{T} \right) (r) \sim \left( \frac{\Delta T}{T} \right) (r_b) \cdot e^{(r-r_b)/H}
\label{eq_DT}
\end{equation}

This exponential divergence should not been confused with the one that is provided by 
Eq.~\ref{eq_sigma_all} for $\sigma_T$.
The latter tends to zero as the noise tends to zero,
while the former is inherent to the nature of Eq.~\ref{eq_n_r}.

\end{document}